\documentclass[twocolumn,prl,aps,superscriptaddress,showpacs]{revtex4}
\usepackage{epsfig}

\begin{document}

\title{Orbital angular momentum driven intrinsic spin Hall effect}

\author{Wonsig Jung}
\affiliation{Institute of Physics and Applied Physics, Yonsei University, Seoul 120-749, Korea}


\author{Dongwook Go}
\affiliation{Department of Physics, Pohang University of Science and Technology, Pohang 790-784, Korea}

\author{Hyun-Woo Lee}
\email{hwl@postech.ac.kr} \affiliation{Department of Physics,
Pohang University of Science and Technology, Pohang 790-784, Korea}

\author{Changyoung Kim}
\email{changyoung@yonsei.ac.kr}
\affiliation{Institute of Physics and Applied Physics, Yonsei University, Seoul 120-749, Korea}

\date{\today}

\begin{abstract}
We propose a mechanism of intrinsic spin Hall effect (SHE). In this mechanism,
local orbital angular momentum (OAM) induces electron position shift and couples with the bias electric field
to generate orbital Hall effect (OHE). SHE then emerges as a concomitant effect of OHE through the atomic spin-orbit coupling.
Spin Hall conductivity due to this mechanism is estimated to be comparable to experimental values for heavy metals.
This mechanism predicts the sign change of the spin Hall conductivity as the spin-orbit polarization changes its sign,
and also correlation between the spin Hall conductivity and the splitting of the Rashba-type spin splitting at surfaces.
\end{abstract}


\maketitle

Spin Hall effect (SHE)~\cite{Hirsch23} is a phenomenon in which electrons with
opposite spins are deflected in opposite side ways.
Its first experimental confirmation~\cite{Kato3} was achieved for
$n$-doped GaAs 
and very small spin
Hall conductivity $\sigma_{SH}\sim 1 \, \, \Omega^{-1}$m$^{-1}$ was
obtained, which was attributed~\cite{Engel2005PRL} to the extrinsic mechanisms~\cite{Zhang24} of SHE
such as skew scattering and side jump.
For some heavy metals, on the other hand,
much larger $\sigma_{SH}$ was reported~\cite{Morota43}. For Pt,
for instance, reported values range from $2.4\times 10^4$
$\Omega^{-1}$m$^{-1}$~\cite{Kimura17} to $5.1\times 10^5$
$\Omega^{-1}$m$^{-1}$~\cite{Ando2008PRL}.
Such large $\sigma_{SH}$ raises hope
for device applications of SHE.
The current-induced magnetization switching observed in Ta/CoFeB magnetic bilayer~\cite{Liu2012Science} is attributed to
the large SHE in Ta, which injects strong spin Hall current into CoFeB to switch its magnetization direction.

Large $\sigma_{SH}$ is often attributed
to intrinsic mechanisms~\cite{Kontani32,Tanaka33,Tanaka34,Guo2008PRL,Guo2005PRL,Murakami25,Sinova26,Kontani2008PRL} of SHE,
which 
do not resort to impurity scattering. Their exact nature remains unclear however.
In one mechanism~\cite{Guo2008PRL},
a small spin-orbit energy gap near the Fermi energy resonantly enhances
the momentum space Berry phase effect to
produce a strong effective magnetic field in {\it momentum} space
and $\sigma_{SH}=10^4 \sim 10^5$ $\Omega^{-1}$m$^{-1}$ is predicted for Pt.
In another mechanism~\cite{Kontani2008PRL,Tanaka34,Tanaka33,Kontani32}, 
the orbital angular momentum (OAM) of atomic orbitals generates the Aharonov-Bohm phase
and produces
a spin-dependent effective magnetic field in {\it real} space.
For various heavy metals with strong atomic spin-orbit (SO) coupling,
resulting $\sigma_{SH}$ is estimated to $10^4 \sim 10^5$ $\Omega^{-1}$m$^{-1}$
and predicted to exhibit a systematic sign change among materials with different spin-orbit polarization,
in qualitative agreement with experiments~\cite{Morota43}.

We report another intrinsic mechanism of SHE
based on a special role of OAM with regard to electron position,
which was not recognized in previous studies~\cite{Kontani2008PRL,Kontani32,Tanaka33,Tanaka34,Bernevig2005PRL} on OAM effect.
For illustration, we use for now a two-dimensional (2D) square lattice in the plane $z=0$.
Later we switch back to 3D.
When $p_z \pm i p_x$ orbitals ($L_y=\pm \hbar$) at different lattice sites
are superposed to form a Bloch state with crystal momentum $\vec{k}$ along $+x$ direction,
the resulting electron density is {\it not} centered around the $z=0$ plane
but instead shifted {\it out-of-plane} along $\pm z$ direction
due to the interference between atomic orbitals at neighboring sites (see Fig.~2 and related discussion in Ref.~\cite{Park37}).
When $\vec{k}$ is small, this shift $\delta \vec{r}$  is given by
\begin{equation}
\delta \vec{r}=\frac{\alpha_K}{e} \vec{k}\times \vec{L},
\label{eq:delta-r}
\end{equation}
for general directions of $\vec{k}$ and $\vec{L}$,
where $-e$ is the electron charge and $\alpha_K$ is a proportionality constant,
which depends on the relative size of atomic orbitals with respect to inter-atomic distance.
Here $\vec{L}$ denotes OAM of {\it atomic} orbitals instead of $\vec{r}\times \hbar \vec{k}$~\cite{Zhang2005PRL}. It thus commutes with $\vec{k}$ and
also with the position operator $\vec{r}$, which is the canonical pair of $\vec{k}$
and measures the lattice position of each atomic orbital.
Nonzero $\delta \vec{r}$ implies that
$\vec{r}$ does not properly represent the true position of an electron.
%
At surfaces with broken inversion symmetry,
this correction couples with an {\it internal} electric field
to produce large Rashba-type spin splitting~\cite{Park37,Kim36}. 

{\it Pedagogical discussion.---}
To illustrate effects of $\delta \vec{r}$
for nonmagnetic systems with inversion symmetry,
we use the free electron-like unperturbed band Hamiltonian $H_0$,
\begin{equation}
H_0=\frac{\hbar^2\vec{k}^2}{2m}+H_{LS}, \label{eq:H_0}
\end{equation}
%
where the atomic SO coupling $H_{LS}$,
\begin{equation}
H_{LS}=\alpha_{SO} \vec{L}\cdot \vec{S},
\end{equation}
is large in heavy metals.
We regard the total angular momentum $J$ as a good quantum number
and illustrate orbital Hall effect (OHE) and SHE for $J=1/2$ states of a 2D electron system.
Note that $H_0$ is two-fold degenerate for all $\vec{k}$ and provides a general description of nonmagnetic systems with
inversion symmetry for small $\vec{k}$.
We remark that for this $H_0$,
previous theories~\cite{Guo2008PRL,Kontani2008PRL,Tanaka33,Kontani32,Tanaka34} of intrinsic SHE do not work.

\begin{figure}
\centering \epsfxsize=8.5cm \epsfbox{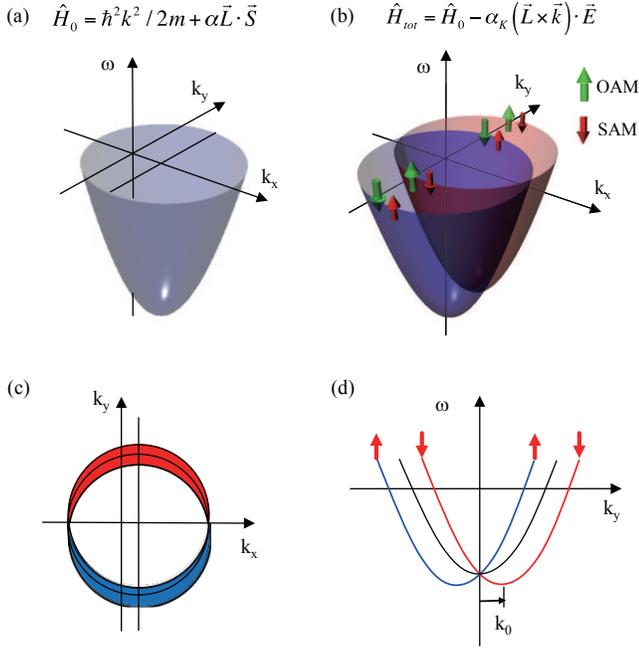}
\caption{(Color
online) Electron dispersion for $J=1/2$ in the
presence of the bias field $\vec{E}$. Band structure based on (a) $H_0$ and
(b) $H_0+H'_2$ together with the occupation change due to $H'_1$.
$\vec{E}$ is applied in the
$-x$-direction. Average spin direction of the split $J_z=\pm 1/2$ bands is anti-parallel to the average OAM direction.
(c) Fermi surfaces of the split bands. Red (blue) area represents
occupied states with only down (up) spins. (d) The band dispersion
along the dash-dot line in (c). $k_0$ is the shift of each band
along the $k_y$ direction.} \label{fig1}
\end{figure}

Our theory deviates from previous theories when a constant external electric field $\vec{E}$ is applied.
The coupling to $\vec{E}$ is commonly given by
\begin{equation}
H'_1=e\vec{E}\cdot \vec{r}.
\end{equation}
However $\delta \vec{r}$ implies that the correct coupling~\cite{Park37} should be $H'_1+H'_2$,
where
\begin{equation}
H'_2=e\vec{E}\cdot \delta \vec{r} = \alpha_K \vec{E}\cdot (\vec{k}\times \vec{L}).
\end{equation}
Previous analyses~\cite{Kontani32,Kontani2008PRL,Tanaka33,Tanaka34} of OAM based intrinsic SHE did not take into account $H'_2$.
%
Thus the total Hamiltonian becomes
\begin{equation}
H_{tot}=H_0+H'_1+H'_2. \label{eq:Total-Hamiltonian}
\end{equation}
Its band, spin angular momentum (SAM), and OAM structures
are plotted for $\vec{E}=-E_{0}\vec{x}$ with $H'_2$ neglected
[Fig.~\ref{fig1}(a)] and with $H'_2$ considered [Fig.~\ref{fig1}(b)]. In addition to
the overall band structure shift in the $k_{x}$-direction as shown
in Fig.~\ref{fig1}(a) (to be more exact, it is actually a shift in the
occupation), the originally degenerate
$J_z=\pm 1/2$ bands get split due to $H'_2$ with the average OAM polarized along the $+z$- or
$-z$-directions as shown in Fig.~\ref{fig1}(b) (exaggerated for a better
view).
The split Fermi surfaces are shown in Fig.~\ref{fig1}(c),
where the Fermi surfaces with opposite OAM are
shifted along opposite $k_{y}$ directions. Consequently, there are
$k$-space regions (shaded areas) where electrons have net OAM;
more electrons with up-OAM in the $+k_{y}$ region
(shaded red) and more electrons with
down-OAM (shaded blue) in the $-k_{y}$ region.
This naturally leads to OHE.
This mechanism of OHE due to $\delta\vec{r}$ differs from other mechanisms~\cite{Bernevig2005PRL,Kontani2008PRL,Tanaka33,Tanaka34,Kontani32} of OHE.

\begin{figure}
\centering \epsfxsize=6.0cm \epsfbox{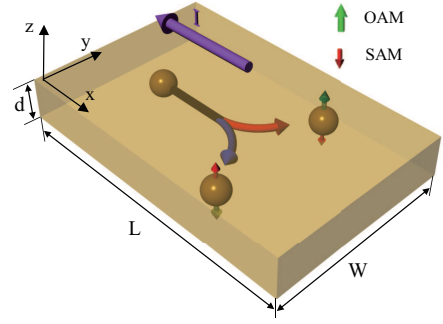}
\caption{(Color
online) Schematic for OAM driven intrinsic SHE. Electrons flow in the $x$-direction by
$\vec{E}$
and are deflected in side ways due to $H'_2$.
Note that the deflection direction depends on the direction of OAM,
amounting to OHE.
For $J=1/2$ band,
$H_{LS}$ sets SAM anti-parallel to OAM.
Thus SHE arises a concomitant effect of OHE.}
\label{fig2}
\end{figure}

For strong $H_{LS}$, OHE implies SHE
since OAM and SAM are correlated;
for $J=1/2$ with $L=1$, they are anti-parallel.
Thus the orbital Hall current implies the spin Hall current of opposite sign.
Figure~\ref{fig2} illustrates the OAM driven intrinsic SHE for the $J=1/2$ case.
This mechanism of SHE can be generalized to other situations in a straightforward way.
For instance, if we apply $H_{tot}$ to the $J=3/2$ case with $L=1$~\cite{Comment-J=3/2-splitting},
one again finds both OHE and SHE, the only qualitative difference 
being that
the orbital and spin Hall currents now have the {\it same} sign
since $\vec{L}\cdot\vec{S}>0$.
This provides an alternative~\cite{Kontani2008PRL,Kontani32,Tanaka33,Tanaka34} explanation for opposite signs of $\sigma_{SH}$
for materials with opposite signs of the SO polarization $\vec{L}\cdot \vec{S}$.

\begin{figure}
\centering \epsfxsize=8.0cm \epsfbox{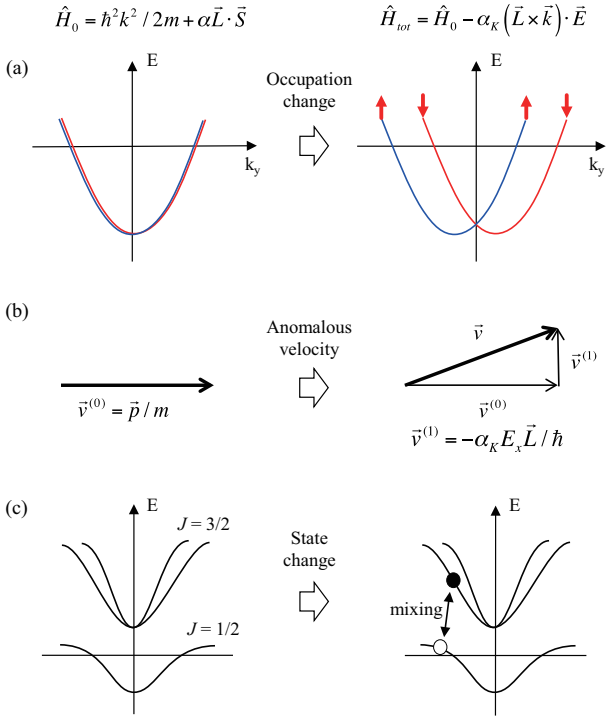}
\caption{(Color online) Schematic illustration of the three terms to the intrinsic SHE, (a) occupation change, (b) anomalous velocity, and (c) state change. Figures on the left represent the situation with $H_0$ while on the right with $H_0+H'_2$.} \label{fig3}
\end{figure}

{\it Conventional spin current.---} The above discussion is incomplete
since it demonstrates only the Fermi surface contribution to SHE
and neglects a Fermi sea contribution.
From now on, we consider a 3D system described by
Eq.~(\ref{eq:Total-Hamiltonian}), and evaluate systematically the conventional spin current density
operator $\hat{j}^{S}_{\alpha,\beta}$ defined by
\begin{equation}
\hat{j}^{S}_{\alpha,\beta}=\frac{1}{V} \frac{-e}{\hbar/2} \frac{ \left\{ S_\alpha, v_\beta \right\}}{2},
\label{eq:Spin-current-density-operator}
\end{equation}
where $\{ \cdots \}$ is the anti-commutator, $V$ is the volume of the system, and the factor $-e/(\hbar/2)$ is introduced to make
$\hat{j}^{S}_{\alpha,\beta}$ have the same dimension as the charge current density.
Here $v_\beta$ is the $\beta$ ($=x,y,z$) component of the velocity operator $\vec{v}$,
\begin{equation}
\vec{v} = \frac{[\vec{r},H_{tot}]}{i\hbar}=\frac{\hbar \vec{k}}{m} + \frac{\alpha_K}{\hbar} (\vec{L}\times \vec{E})=\vec{v}^{(0)}+\vec{v}^{(1)}.
\end{equation}
Note that $\vec{v}$ contains two contributions.
When the anomalous velocity $\vec{v}^{(1)}=(\alpha_K/\hbar) \vec{L}\times \vec{E}$ is neglected
and the resulting $\hat{j}^{S}_{\alpha,\beta}$ is averaged over the shaded momentum space region in Fig.~\ref{fig1}(c)
[to be precise, 3D counterpart of Fig.~\ref{fig1}(c)],
one obtains what we call the occupation change contribution $(j^{S}_{\alpha,\beta})_{oc}$
coming from the Fermi surface, as illustrated in the pedagogical discussion.
The magnitude of $(j^{S}_{\alpha,\beta})_{oc}$ can be estimated easily.
The density of electrons that contribute to the {\it net} spin current density
is proportional to $4\pi k_F^2 k_0$, where $k_F$ is the Fermi wavevector for the unperturbed Fermi surface
and $k_0\sim (m/\hbar^2)\alpha_K \vec{E}\times \vec{L}$
is the Fermi surface shift caused by $H'_2$ [see Fig.~\ref{fig1}(d)].
Each of such electrons contributes $\pm e$ for $[-e/(\hbar/2)]S_\alpha$, and
$\pm \hbar k_F/m$ for $v_\beta$.
Combined with a symmetry consideration, which requires $(j^{S}_{\alpha,\beta})$ to be proportional to $\epsilon_{\alpha\beta\gamma}E_\gamma$,
where $\epsilon_{\alpha\beta\gamma}$ is the Levi-Civita symbol, one finds
\begin{equation}
(j^{S}_{\alpha,\beta})_{oc}=(\eta_J)_{oc}\epsilon_{\alpha\beta\gamma}E_\gamma e \alpha_K \frac{4\pi k_F^3/3}{(2\pi)^3},
\end{equation}
where $(\eta_J)_{oc}$ is a dimensionless constant.
From the exact evaluation~\cite{SupplMaterial} of $(j^{S}_{\alpha,\beta})_{oc}$,
we find
$(\eta_{J=1/2})_{oc}=4/9$ and $(\eta_{J=3/2})_{oc}=-20/9$~\cite{Comment-J=3/2-splitting}.
Note that the sign of $\eta_J$ is opposite for the two $J$'s as expected.
%

The anomalous velocity $v^{(1)}$ generates additional contribution, which comes from the Fermi sea.
When $v^{(0)}$ is neglected and only $v^{(1)}$ is retained,
the average of the resulting $\hat{j}^{S}_{\alpha,\beta}$ over the unperturbed Fermi sea of $H_0$
results in what we call the anomalous velocity contribution $(j^{S}_{\alpha,\beta})_{av}$.
To estimate its magnitude, one first notes that
$\epsilon_{\alpha\beta\gamma}S_\alpha v^{(1)}_\beta = (\alpha_K/\hbar)[\vec{S}\times (\vec{L}\times \vec{E})]_\gamma
=(\alpha_K/\hbar) [(\vec{S}\cdot \vec{E})\vec{L}-(\vec{S}\cdot \vec{L})\vec{E}]_\gamma$.
While the first term may fluctuate in sign,
the second term ($\propto \vec{S}\cdot\vec{L}$) has a definite sign over the Fermi sea.
Thus $(j^{S}_{\alpha,\beta})_{av}$ may be estimated by multiplying the second term with
the electron density $\sim (4\pi k_F^3/3)/(2\pi)^3$, which results in
\begin{equation}
(j^{S}_{\alpha,\beta})_{av}=(\eta_J)_{av}\epsilon_{\alpha\beta\gamma}E_\gamma e \alpha_K \frac{4\pi k_F^3/3}{(2\pi)^3},
\end{equation}
where $(\eta_J)_{av}$ is a dimensionless constant.
From the exact evaluation~\cite{SupplMaterial} of $(j^{S}_{\alpha,\beta})_{av}$,
we find
$(\eta_{J=1/2})_{av}=-4/3$ and $(\eta_{J=3/2})_{av}=+4/3$~\cite{Comment-J=3/2-splitting}.
The sign of $(\eta_J)_{av}$ is again opposite for the two $J$ values
due to the sign difference of $\vec{S}\cdot\vec{L}$.
%
Figures~\ref{fig3}(a) and (b) illustrate schematically $(j^{S}_{\alpha,\beta})_{oc}$
and $(j^{S}_{\alpha,\beta})_{av}$.

In addition, there exists a third contribution which is illustrated in Fig.~\ref{fig3}(c).
When $\vec{E}$ is applied, $\vec{J}=\vec{L}+\vec{S}$ is not a good quantum number any more
and $H'_2$ induces the inter-band mixing between the $J=1/2$ and $J=3/2$ bands.
This contribution $(j^{S}_{\alpha,\beta})_{sc}$, which we call the state change contribution,
is inversely proportional to the band separation between the $J=1/2$ and $J=3/2$ bands,
and becomes smaller as $H_{LS}$ becomes larger.
From the exact evaluation~\cite{SupplMaterial} of $(j^{S}_{\alpha,\beta})_{sc}$,
we find that $(j^{S}_{\alpha,\beta})_{sc}$ is smaller than $(j^{S}_{\alpha,\beta})_{oc}$ and $(j^{S}_{\alpha,\beta})_{av}$
by the factor $(\hbar^2 k_F^2/2m)/\Delta E$,
where $\Delta E=3\hbar^2 \alpha_{SO}/2$ is the energy separation between the $J=1/2$ and $J=3/2$ bands.
Since we are interested in the large $H_{LS}$ limit, we ignore $(j^{S}_{\alpha,\beta})_{sc}$ in the subsequent discussion.

{\it Proper spin current.---}
Next we examine whether the OAM driven spin Hall current generates spin accumulation at side surfaces of a system,
which is what is actually measured in SHE detection schemes
such as Kerr rotation spectroscopy~\cite{Kato3,Sih6,Sih7,Stern8,Stern11,Matsusaka12}
and photoluminescence~\cite{Wunderlich4,Chang9}.
Since $H_{LS}$ breaks the spin conservation, nonzero conventional spin current does not guarantee the spin accumulation~\cite{Murakami2004PRL}.
For transparent connection with the spin accumulation, we evaluate the proper spin current density operator~\cite{Shi2006PRL}
\begin{equation}
\hat{j}^{S,prop}_{\alpha,\beta}=\frac{1}{V} \frac{-e}{\hbar/2} \frac{ d( S_\alpha r_\beta) }{dt},
\end{equation}
%
which captures the {\it combined} effect of the conventional spin current and the spin conservation violation.
We evaluate~\cite{SupplMaterial} the spin current for $H_{tot}$ by using $\hat{j}^{S,prop}_{\alpha,\beta}$
instead of $\hat{j}^{S}_{\alpha,\beta}$ [Eq.~(\ref{eq:Spin-current-density-operator})],
and find identical results,
confirming the spin accumulation by the OAM driven SHE.

To be more rigorous, however, both $\hat{j}^{S,prop}_{\alpha,\beta}$ and $\hat{j}^{S}_{\alpha,\beta}$ 
fail to capture the full effect of $\delta \vec{r}$,
since both operators are defined in terms of $\vec{r}$, which does not represent the true position of electrons.
To remedy this problem, $\vec{r}$ in the definitions should be replaced by $\vec{R}\equiv \vec{r} +\delta \vec{r}$.
After this remedy to $\hat{j}^{S,prop}_{\alpha,\beta}$, we find~\cite{SupplMaterial} that
the anomalous contribution $(j^S_{\alpha,\beta})_{av}$ becomes doubled.
Thus in the strong $H_{LS}$ limit, the total spin Hall conductivity $\sigma_{SH}$ [$(j^{S}_{\alpha,\beta})_{total}=\epsilon_{\alpha\beta\gamma}\sigma_{SH}E_\gamma]$ is given by
\begin{equation}
\sigma_{SH}=(\eta_J)_{total} e \alpha_K \frac{4\pi k_F^3/3}{(2\pi)^3}
\label{eq:SH-conductivity}
\end{equation}
for small $\vec{k}$, where the dimensionless constant $(\eta_J)_{total}=(\eta_J)_{oc}+2(\eta_J)_{av}$
is $4/9-8/3=-20/9$ for $J=1/2$ and $-20/9+8/3=4/9$ for $J=3/2$.
Note that $\sigma_{SH}$ has opposite signs for $J=1/2$ and $J=3/2$.

{\it Discussion.---}
To understand better the mechanism of the OAM driven SHE, it is useful to
examine the equation of motion,
$d\vec{R}/dt=\vec{v}^{(0)}+2\vec{v}^{(1)}-(\alpha_K/e)\vec{k}\times d\vec{S}/dt+[\delta \vec{r},e\vec{E}\cdot \delta \vec{r}]/i\hbar$.
The factor 2 in the second term explains why $(j^S_{\alpha,\beta})_{av}$ is doubled
after the remedy to $\hat{j}^{S,prop}_{\alpha,\beta}$. 
The third term vanishes in the steady state
and does not contribute to $\sigma_{SH}$~\cite{SupplMaterial}.
The last term ($\propto k^2$) is small in the small $\vec{k}$ limit but is important conceptually.
Further insights can be gained by regarding $\delta \vec{r}$ as a momentum space vector potential $\vec{A}\equiv -\delta \vec{r}$.
Then $\vec{R}=\vec{r}-\vec{A}$ amounts to the ``gauge-invariant" position operator.
The equations of motion become
\begin{eqnarray}
\frac{d\vec{R}}{dt}= \frac{\hbar \vec{k}}{m}
+ \frac{d\vec{k}}{dt}\times \vec{B}, \ \
\frac{d\vec{k}}{dt} = -\frac{e\vec{E}}{\hbar},
\label{eq:equation-of-motion}
\end{eqnarray}
where the momentum space effective magnetic field $B_\alpha=(1/2)\epsilon_{\alpha\beta\gamma}{\cal F}_{\beta\gamma}$ with
%
\begin{equation}
{\cal F}_{\beta\gamma}=\partial_{k_\beta}A_\gamma - \partial_{k_\gamma}A_\beta + i [A_\beta,A_\gamma].
\label{eq:Field-strength}
\end{equation}
Note that Eq.~(\ref{eq:equation-of-motion}) has the same form as the wavepacket equations of motion~\cite{Sundaram1999PRB}
in the presence of the momentum space Berry phase.
There is however an important difference; the momentum space Berry connection $\vec{A}$ is now non-Abelian
($[A_\beta,A_\gamma]=[\delta r_\beta,\delta r_\gamma]\neq 0$).
For the non-Abelian case, the commutator in Eq.~(\ref{eq:Field-strength}) is crucial
to keep the field strength tensor ${\cal F}_{\beta\gamma}$ ``gauge-invariant".
This indicates that $\delta \vec{r}$ induces
the momentum space non-Abelian Berry phase, which is responsible
for the Fermi sea contribution $2(\eta_J)_{av}$ to $\sigma_{SH}$.
The non-Abelian $\vec{A}$ also implies the noncommutative space, $[R_\alpha,R_\beta]\neq 0$.
Such noncommutative geometry arises generically when the true position operator is projected
onto a sub-Hilbert space~\cite{Shindou2005NP}. A well known example is the quantum Hall effect,
where the noncommutativity emerges after the projection onto the lowest Landau level~\cite{Polychronokos2000JHEP}.
For the present case, the noncommutativity arises since $\vec{R}$ amounts to the projection of the true position operator
onto the sub-Hilbert space with fixed $\vec{L}^2$ ($L=1$).

Difference from other mechanisms of intrinsic SHE is now evident.
Unlike previous works on the OAM based SHE~\cite{Kontani32,Kontani2008PRL,Tanaka33,Tanaka34},
what OAM generates is the momentum space Berry phase
instead of the real space Aharonov-Bohm phase.
Unlike previous works~\cite{Guo2008PRL} based on the momentum space Berry phase,
its origin is $\delta \vec{r}$ instead of small SO gap.
Thus this mechanism works even for $J=1/2$,
for which SO gap is forbidden in nonmagnetic systems with inversion symmetry.
In this sense, this mechanism is quite generic; it applies to all non-$s$-character orbitals,
with the only serious constraint being large $H_{LS}$.
When $H_{LS}$ is small, bands with opposite signs of $\vec{L}\cdot\vec{S}$ overlap
and their contributions to $\sigma_{SH}$ tend to cancel each other.

Finally we estimate the magnitude of $\sigma_{SH}\sim e \alpha_K n$, where $n$ is the electron density.
To estimate $\alpha_K$, we utilize the connection between $\alpha_K$ and the Rashba-type SO coupling constant $\alpha_{R}$
near a surface where the structural inversion symmetry is broken.
Some of us have demonstrated~\cite{Park37,Kim36} that
the maximum $\alpha_R$ in the large $H_{LS}$ limit
is roughly given by $\alpha_K |\vec{E}_{int}|\hbar$, 
where $\vec{E}_{int}$ denotes the internal electric field near surfaces produced by the inversion symmetry breaking
and is of order (work function)/(atomic spacing) $\sim 1$ V/A.
For $\alpha_R\sim 10^{-11}-10^{-10}$ eV$\cdot$m~\cite{Varykhalov2008PRL,Bendounan2011PRB,Park2013PRL},
one obtains $\alpha_K\sim 10^{-6}-10^{-5}$ m$^2$V$^{-1}$s$^{-1}$.
Then for typical metallic electron density $n\sim (3\ {\rm A})^{-3}$,
one obtains $\sigma_{SH}\sim 10^4 - 10^5 \ \Omega^{-1}$m$^{-1}$, which is comparable to experimental values for heavy metals~\cite{Morota43}.
We note however that this estimation is crude
since Eq.~(\ref{eq:SH-conductivity}) is derived in the small $\vec{k}$ limit
whereas $\vec{k}$ is not small in metallic systems.
Moreover it ignores complicated band structures of real materials.

In conclusion, we presented a generic mechanism of intrinsic SHE based on OAM,
which is applicable to all non-$s$-character orbitals in nonmagnetic systems with inversion symmetry.
The position shift $\delta \vec{r}$ due to OAM
gives rise to the non-Abelian Berry curvature in the momentum space,
which produces both OHE and SHE.
This mechanism
implies the sign change of $\sigma_{SH}$ as the SO polarization $\vec{S}\cdot \vec{L}$ changes its sign.
The resulting $\sigma_{SH}$ is estimated to $10^4-10^5\ \Omega^{-1}$m$^{-1}$ when $H_{LS}$ is large.
This OAM based theory also predicts the correlation between $\sigma_{SH}$ and the strength of the Rashba-type spin splitting at surfaces.

We acknowledge fruitful discussion with G. S. Jeon, J. H. Han, K. J. Lee and B. C. Min.
This research was supported by the Converging Research Center Program through the Ministry of Science, ICT and Future Planning, Korea(2013K000312). HWL acknowledges the financial support of the NRF (2011-0030784 and 2013R1A2A2A05006237).

\end{document}


\renewcommand{\theequation}{S\arabic{equation}}
\renewcommand{\thefigure}{S\arabic{figure}}
\title{Supplementary Material: Orbital angular momentum driven intrinsic spin Hall effect}

\author{W. S. Jung}
\affiliation{Institute of Physics and Applied Physics, Yonsei University, Seoul 120-749, Korea}

\author{Dongwook Go}
\affiliation{Department of Physics,
Pohang University of Science and Technology, Pohang, Kyungbuk
790-784, Korea}

\author{Hyun-Woo Lee}
\email{hwl@postech.ac.kr} \affiliation{Department of Physics,
Pohang University of Science and Technology, Pohang, Kyungbuk
790-784, Korea}

\author{C. Kim}
\email{changyoung@yonsei.ac.kr}
\affiliation{Institute of Physics and Applied Physics, Yonsei University, Seoul 120-749, Korea}

\date{\today}

\maketitle

In Secs.~\ref{sec1}, \ref{sec2}, \ref{sec3} of the supplementary
material, we present the calculation of the spin current density
in 3D. Eventually we calculate in Sec.~\ref{sec3} the proper spin
current density $j_{\alpha,\beta}^{S,PROP}$, which is based on the
concept of the ``proper'' spin current~\cite{Shi2006PRL} and formulated in
terms of the ``proper" position operator $\vec{R}$. However its
calculation is rather technical and less illuminating. Thus for
pedagogical purpose, we present the calculation of more
conventional spin current density first in Secs.~\ref{sec1} and
\ref{sec2}.
%

In Sec.~\ref{sec1}, we present the calculation of the conventional
spin current density $j_{\alpha,\beta}^{S}$ formulated in terms of
the conventional position operator $\vec{r}$, where $\vec{r}$ is
the canonical pair of the Bloch momentum $\vec{k}$ and
$j_{\alpha,\beta}^{S}$ is the expectation value of the
conventional spin current density operator
$\hat{j}_{\alpha,\beta}^{S}$,
\begin{equation}
\hat{j}_{\alpha,\beta}^{S}=\frac{1}{V}\frac{-e}{\hbar/2}\frac{ \{
S_\alpha, dr_\beta/dt \} }{2}.
\label{eq:Conventional-spin-current-operator}
\end{equation}
Note that $\hat{j}^{S}_{\alpha,\beta}$ is defined to have the same
dimension as the charge current density. We demonstrate that
$j_{\alpha,\beta}^{S}$ has three independent contributions, which
we call the anomalous velocity contribution, the state change
contribution, and the occupation change contribution. The physical
meaning of each contribution will become clear in Sec.~\ref{sec1}.

In Sec.~\ref{sec2}, we present the calculation of the proper spin
current density $j_{\alpha,\beta}^{S,prop}$ formulated in terms of
the conventional position operator $\vec{r}$. The concept of the
proper spin current was proposed~\cite{Shi2006PRL} to take into
account the violation of the spin conservation and to facilitate
the connection with the spin accumulation.
$j_{\alpha,\beta}^{S,prop}$ is the expectation value of the
operator $\hat{j}_{\alpha,\beta}^{S,prop}$,
\begin{equation}
\hat{j}_{\alpha,\beta}^{S,prop}=\frac{1}{V}\frac{-e}{\hbar/2}\frac{d}{dt}\frac{ \{ S_\alpha, r_\beta \} }{2}.
\label{eq:Proper-spin-current-operator}
\end{equation}
Compared to Eq.~(\ref{eq:Conventional-spin-current-operator}),
where the time derivative applies to $r_\beta$ only,
Eq.~(\ref{eq:Proper-spin-current-operator}) differs since the time
derivative now applies to the anti-commutator
$\{S_\alpha,r_\beta\}$. We demonstrate that
$j_{\alpha,\beta}^{S,prop}$ is identical to
$j_{\alpha,\beta}^{S}$.

In Sec.~\ref{sec3}, we finally present the calculation of the
proper spin current density $j_{\alpha,\beta}^{S,PROP}$ formulated
in terms of the proper position operator $\vec{R}$, where
$\vec{R}$ differs from $\vec{r}$ as follows,
\begin{equation}
\vec{R}=\vec{r}+\frac{\alpha_K}{e}\vec{k}\times \vec{L},
\label{eq:Proper-position-operator}
\end{equation}
and $j_{\alpha,\beta}^{S,PROP}$ is the expectation value of the
operator $\hat{j}_{\alpha,\beta}^{S,PROP}$,
\begin{equation}
\hat{j}_{\alpha,\beta}^{S,PROP}=\frac{1}{V}\frac{-e}{\hbar/2}\frac{d}{dt}\frac{
\{ S_\alpha, R_\beta \} }{2}.
\label{eq:Consistent-spin-current-operator}
\end{equation}
Note that Eq.~(\ref{eq:Consistent-spin-current-operator}) is
identical to Eq.~(\ref{eq:Proper-spin-current-operator}) except
that $R_\beta$ appears instead of $r_\beta$. While the calculation
of $j_{\alpha,\beta}^{S,PROP}$ is more tedious than those of the
former two counterparts, the value of $j_{\alpha,\beta}^{S,PROP}$
turns out to be almost identical to $j_{\alpha,\beta}^{S}$ and
$j_{\alpha,\beta}^{S,prop}$, except that the magnitude of the
anomalous velocity contribution is now two times bigger.

\section{Conventional spin current density}
\label{sec1}
Here we present the calculation of the conventional spin current
density $j^{S}_{\alpha,\beta}$ formulated in terms of the
conventional position operator. $j^{S}_{\alpha,\beta}$ is given by
\begin{equation}
j^{S}_{\alpha,\beta}= {\rm Tr}\left[ \hat{j}^{S}_{\alpha,\beta}
\hat{\rho} \right],
\label{eq:Spin-current-density-evaluation-formula}
\end{equation}
where $\hat{\rho}$ is the density matrix and the operator
$\hat{j}^{S}_{\alpha,\beta}$ is defined in
Eq.~(\ref{eq:Conventional-spin-current-operator}).


For $\vec{E}=0$, both $H'_1$ and $H'_2$ vanish and $\hat{\rho}$
becomes its equilibrium form $\hat{\rho}^{(0)}$, where
\begin{equation}
\hat{\rho}^{(0)}= \sum_n
f^{(0)}\!\!\left(E^{(0)}_n\right)\left|n\right\rangle^{\!(0)}
\sideset{^{(0)}\!}{}{\mathop{ \left\langle n \right|}},
\label{eq:Equilbrium-density-matrix}
\end{equation}
Here $|n\rangle^{(0)}$ denotes an eigenstate of $H_0$ with energy
eigenvalue $E^{(0)}_n$, and $f^{(0)}(E)$ is the equilibrium Fermi
occupation function.
It is straightforward to verify that $j^{S}_{\alpha,\beta}$
vanishes in equilibrium.

When a nonzero $\vec{E}$ is applied, we evaluate
$j^{S}_{\alpha,\beta}$ up to the first order in $\vec{E}$. Up to
this order, effects of $H'_1$ and $H'_2$ may be considered
separately. $H'_1$ alone does not contribute to
$j^{S}_{\alpha,\beta}$ at all since as far as $H_0 + H'_1$ is
concerned, the dynamics of $\vec{r}$ in $H'_1$ is decoupled from
that of $\vec{S}$. This is evident from the facts that $H'_1$
commutes with both $\vec{L}$ and $\vec{S}$ and that there is no
coupling in $H_0+H'_1$ linking $\vec{r}$ (or $\vec{k}$) with
$\vec{S}$ (or $\vec{L}$).
Below we thus ignore effects of $H'_1$ and consider effects of
$H'_2$ only.

In Secs.~\ref{sec:j-spin-av} and \ref{sec:j-spin-im}, we evaluate
two contributions to $j^{S}_{\alpha,\beta}$ under the assumption
that impurity scattering is completely absent. In
Sec.~\ref{sec:j-spin-oc}, we consider the effect of the impurity
scattering on $j^{S}_{\alpha,\beta}$ in the limit of vanishingly
weak scatterers.

\subsection{Anomalous velocity contribution}
\label{sec:j-spin-av}
One effect of $H'_2$ is to modify the velocity operator $\vec{v}$.
For the total Hamiltonian $H_0+H'_2$, $\vec{v}$ is given by
\begin{equation}
\vec{v}=\frac{d\vec{r}}{dt}=\frac{[\vec{r},H_0+H'_2]}{i\hbar}=\frac{\hbar \vec{k}}{m}+\frac{\alpha_K}{\hbar} \vec{L}\times \vec{E}=\vec{v}^{(0)}+\vec{v}^{(1)},
\label{eq:Conventional-velocity-operator}
\end{equation}
where $\vec{v}^{(0)}$ and $\vec{v}^{(1)}$ refer to the terms independent of and linear in $\vec{E}$.
Here we call $\vec{v}^{(1)}$ the anomalous velocity since it denotes the extra contribution to the velocity generated by $\vec{E}$.

$\vec{v}^{(1)}$ generates what we call the anomalous velocity
contribution $(j^{S}_{\alpha,\beta})_{av}$ to the spin current,
\begin{equation}
(j^{S}_{\alpha,\beta})_{av}  =  \frac{1}{V}\frac{-e}{\hbar/2} {\rm
Tr}\left[  \frac{ \{ S_\alpha, v^{(1)}_\beta \}}{2} \hat{\rho}
\right]. \label{eq:j-spin-v1}
\end{equation}
Up to the first order in $\vec{E}$, $\hat{\rho}$ in the above equation may be replaced by $\hat{\rho}^{(0)}$ since $v^{(1)}$ is already first order in $\vec{E}$. Then Eq.~(\ref{eq:j-spin-v1}) reduces to
\begin{equation}
(j^{S}_{\alpha,\beta})_{av} = \frac{1}{V}\frac{-e}{\hbar/2}\sum_n
f^{(0)}\!\!\left(E^{(0)}_n\right) \sideset{^{(0)}\!}{}{\mathop{
\left\langle n \right|}} S_\alpha v^{(1)}_\beta \left| n
\right\rangle^{\!(0)}. \label{eq:j-spin-av}
\end{equation}
Here one used $S_\alpha v^{(1)}_\beta = v^{(1)}_\beta S_\alpha$.
Since the eigenstates of $H_0$ are completely specified by the three quantum numbers ($\vec{k}$, $J$, $J_z$) within the orbital angular momentum $L=1$ sector,
the state $|n\rangle^{(0)}$ amounts to $|\vec{k}, J, J_z\rangle^{(0)}$.
For a given $J$, the summation over $n$ in Eq.~(\ref{eq:j-spin-av}) amounts to the summations over $\vec{k}$ and $J_z$.
Since $E^{(0)}_n=E^{(0)}(\vec{k},J)$ is independent of $J_z$,
the summation over $J_z$ leads to the following partial trace over $j_z$,
\begin{equation}
\sum_{J_z} \sideset{^{(0)}\!\!}{}{\mathop{ \left\langle \vec{k},J,J_z \right|}}S_\alpha v^{(1)}_\beta \left| \vec{k},J,J_z \right\rangle^{\!\!(0)}.
\end{equation}
One then utilizes the relations $v^{(1)}_\beta = (\alpha_K/\hbar) \epsilon_{\beta \eta \gamma} L_\eta E_\gamma$ and
\begin{eqnarray}
&& \sum_{J_z} \sideset{^{(0)}\!\!}{}{\mathop{ \left\langle \vec{k},J,J_z \right|}}S_\alpha L_\eta \left| \vec{k},J,J_z \right\rangle^{\!\!(0)} \\
&=& \delta_{\alpha,\eta} \sum_{J_z} \sideset{^{(0)}\!\!}{}{\mathop{ \left\langle \vec{k},J,J_z \right|}}S_z L_z \left| \vec{k},J,J_z \right\rangle^{\!\!(0)} \nonumber \\
&=& \frac{1}{3} \delta_{\alpha\eta} \sum_{J_z} \sideset{^{(0)\!\!}}{}{\mathop{ \left\langle \vec{k},J,J_z \right|}}\vec{S}\cdot \vec{L} \left| \vec{k},J,J_z \right\rangle^{\!\!(0)}. \nonumber
\end{eqnarray}
Note that $\vec{S}\cdot \vec{L}=(\hbar^2/2)\left[ J(J+1) - 1 \cdot
2 - \frac{1}{2}\cdot \frac{3}{2} \right]$ has opposite signs for
$J=3/2$ and $J=1/2$. As confirmed below, this sign difference
leads to the sign difference in $(j^{S}_{\alpha,\beta})_{av}$ for
$J=3/2$ and $J=1/2$. Subsequent calculation proceeds as follows.
One first obtains
\begin{eqnarray}
&& \sum_{J_z} \sideset{^{(0)\!\!}}{}{\mathop{ \left\langle \vec{k},J,J_z \right|}}S_\alpha v^{(1)}_\beta \left| \vec{k},J,J_z \right\rangle^{\!\!(0)} \\
&=& -\epsilon_{\alpha\beta\gamma}  E_\gamma \frac{\alpha_K}{\hbar}\hbar^2\frac{1}{6} \left[J(J+1) - \frac{11}{4} \right](2J+1). \nonumber
\end{eqnarray}
Then the anomalous velocity contribution to the spin current density becomes
\begin{eqnarray}
&& (j^{S}_{\alpha,\beta})_{av}
\label{eq:j-conv-anomalous-final-result} \\
&=& \frac{-e}{\hbar/2} (-\epsilon_{\alpha\beta\gamma}) \frac{\alpha_K}{\hbar} E_\gamma \hbar^2\frac{1}{6} \left[J(J+1) - \frac{11}{4} \right] (2J+1) \nonumber \\
&& \times \frac{1}{V}\sum_{\vec{k}}f^{(0)}\left(E^{(0)}(\vec{k},J)\right) \nonumber \\
&=& \epsilon_{\alpha\beta\gamma}E_\gamma e\alpha_K \frac{1}{3} \left[J(J+1) - \frac{11}{4} \right] (2J+1) \frac{4\pi k_F^3/3}{(2\pi)^3} \nonumber \\
&=& \pm \frac{2}{9\pi^2} \epsilon_{\alpha\beta\gamma}E_\gamma e\alpha_K k_F^3, \nonumber
\end{eqnarray}
where the upper and lower signs apply to $J=3/2$ and $J=1/2$,
respectively. Note that $(j^{S}_{\alpha,\beta})_{av}$ indeed has
opposite signs for $J=3/2$ and $J=1/2$. This sign difference stems
from the fact that $\vec{S}$ is parallel (antiparallel) to
$\vec{L}$ for $J=3/2$ ($J=1/2$).

\subsection{State change contribution}
\label{sec:j-spin-im}
In addition to Eq.~(\ref{eq:j-spin-v1}), which captures the effect of the anomalous velocity $\vec{v}^{(1)}$,
the conventional velocity operator $\vec{v}^{(0)}$ also contributes to the spin current density.
We call this contribution the state change contribution for the reason that will become clear below.
It is given by
\begin{equation}
(j^{S}_{\alpha,\beta})_{sc}=\frac{1}{V}\frac{-e}{\hbar/2} {\rm
Tr}\left[  \frac{ \{ S_\alpha, v^{(0)}_\beta \}}{2} \hat{\rho}
\right]. \label{eq:j-spin-v0}
\end{equation}
When $\hat{\rho}$ in the above expression is replaced by
$\hat{\rho}^{(0)}$, the above expression vanishes. Thus
$(j^{S}_{\alpha,\beta})_{sc}$ arises from the first order
correction to $\hat{\rho}$ due to $\vec{E}$. Up to this order, one
obtains
\begin{equation}
(j^{S}_{\alpha,\beta})_{sc}=\frac{1}{V}\frac{-e}{\hbar/2} {\rm
Tr}\left[  S_\alpha v^{(0)}_\beta \hat{\rho}^{(1)} \right],
\label{eq:j-spin-im}
\end{equation}
where $S_\alpha v^{(0)}_\beta =  v^{(0)}_\beta S_\alpha$ is used.
One way to evaluate Eq.~(\ref{eq:j-spin-im}) is to use the Kubo
formula. Here we evaluate Eq.~(\ref{eq:j-spin-im}) in a slightly
different way, since this alternative method illustrates better
why $(j^{S}_{\alpha,\beta})_{sc}$ may be called the inter-band
mixing contribution. It is straightforward to verify that this
method and the Kubo formula produce the same result for
$(j^{S}_{\alpha,\beta})_{sc}$.

The adiabatic turning-on procedure allows a straightforward evaluation of $\hat{\rho}^{(1)}$.
When $H'_2$ is turned on adiabatically from the far past $t=-\infty$, $\hat{\rho}$ at present time $t=0$ is given by
\begin{equation}
\hat{\rho}= \sum_n
f^{(0)}\!\!\left(E^{(0)}_n\right)\left|n\right\rangle
\sideset{}{}{\mathop{ \left\langle n \right|}}.
\end{equation}
Here $|n\rangle$ denotes the state at $t=0$, to which $|n\rangle^{(0)}$ at $t=-\infty$ evolves  as $H'_2$ is adiabatically turned on.
Up to the first order in $\vec{E}$, $|n\rangle$ differs from $|n\rangle^{(0)}$ by $|n\rangle^{(1)}$, which is given by
\begin{equation}
\left| n \right\rangle^{\!(1)}=\sum_{n'\neq n}|n'\rangle^{(0)} \frac{ \sideset{^{(0)}\!}{}{\mathop{\langle n'|}} H'_2 |n\rangle^{(0)}}{E^{(0)}_n - E^{(0)}_{n'}}.
\label{eq:n-1}
\end{equation}
Then $\hat{\rho}^{(1)}$ becomes
\begin{equation}
\hat{\rho}^{(1)}=\sum_n
f^{(0)}\!\!\left(E^{(0)}_n\right)
\left( \left|n\right\rangle^{\!(0)}
\sideset{^{(1)}\!}{}{\mathop{ \left\langle n \right|}}
+\left|n\right\rangle^{\!(1)}
\sideset{^{(0)}\!}{}{\mathop{ \left\langle n \right|}} \right),
\label{eq:density-matrix-no-scattering}
\end{equation}
and $(j^{S}_{\alpha,\beta})_{sc}$ in Eq.~(\ref{eq:j-spin-im})
becomes
\begin{eqnarray}
(j^{S}_{\alpha,\beta})_{sc} &=& \frac{1}{V}\frac{-e}{\hbar/2}
\sum_n \sum_{n'} f^{(0)}\!\!\left(E^{(0)}_{n}\right)
\label{eq:j-spin-im-1} \\
&& \times \left(
\sideset{^{(0)}\!}{}{\mathop{ \left\langle n' \right|}} S_\alpha v^{(0)}_\beta \left| n \right\rangle^{\!(0)}
\sideset{^{(1)}\!}{}{\mathop{ \left\langle n \right|}} \left. n' \right\rangle^{\!(0)} \right. \nonumber \\
&& + \left.
\sideset{^{(0)}\!}{}{\mathop{ \left\langle n' \right|}} S_\alpha v^{(0)}_\beta \left| n \right\rangle^{\!(1)}
\sideset{^{(0)}\!}{}{\mathop{ \left\langle n \right|}} \left. n' \right\rangle^{\!(0)} \right). \nonumber
\end{eqnarray}

To evaluate this expression,
one recalls $E^{(0)}_n$ being independent of $J_z$ and exploits this energy degeneracy to introduce a new set of quantum numbers
($\vec{k}$, $J$, $J_{\tilde{z}}$) to specify the state $n$. Here $J_{\tilde{z}}$ denotes the component of the total angular momentum along
the direction $\tilde{z}$, which points along $\vec{E}\times \vec{k}$ direction. Note that $\tilde{z}$ axis is dependent on $\vec{k}$.
This change of the angular momentum quantization axis from $z$ to $\tilde{z}$ simplifies the evaluation of Eq.~(\ref{eq:n-1}).
Considering that $H'_2$ reduces to $\alpha_K |\vec{E}\times\vec{k}| L_{\tilde{z}}$, one finds
\begin{eqnarray}
&& \sideset{^{(0)}\!}{}{\mathop{\left\langle n'\right|}} H'_2 \left|n\right\rangle^{\!(0)} \\
&=& \alpha_K |\vec{E}\times \vec{k}|  \sideset{^{(0)}\!\!}{}{\mathop{\left\langle \vec{k}',J',J'_{\tilde{z}} \right|}} L_{\tilde{z}} \left|\vec{k},J,J_{\tilde{z}} \right\rangle^{\!\!(0)}  \nonumber \\
&=& \delta_{\vec{k}'\vec{k}}\delta_{J'_{\tilde{z}}J_{\tilde{z}}}
\alpha_K |\vec{E}\times \vec{k}|  \sideset{^{(0)}\!\!}{}{\mathop{\left\langle \vec{k},J',J_{\tilde{z}} \right|}} L_{\tilde{z}}
\left|\vec{k},J,J_{\tilde{z}} \right\rangle^{\!\!(0)}.  \nonumber
\end{eqnarray}
Thus $H'_2$ induces the inter-band mixing between $|\vec{k},J=1/2,J_{\tilde{z}}\rangle^{(0)}$ and $|\vec{k},J=3/2,J_{\tilde{z}}\rangle^{(0)}$.
It is now evident that $(j^{S}_{\alpha\beta})_{sc}$  captures the
effect of the state change due to the inter-band mixing caused by
$H'_2$.
The matrix elements that capture this inter-band mixing effect are
\begin{eqnarray}
&& \sideset{^{(0)}\!\!}{}{\mathop{\left\langle \vec{k},J=\frac{1}{2},J_{\tilde{z}}=\pm \frac{1}{2} \right|}}H'_2 \left| \vec{k},J=\frac{3}{2},J_{\tilde{z}}=\pm\frac{1}{2} \right\rangle^{\!\!(0)}  \nonumber\\
&=&  \sideset{^{(0)}\!\!}{}{\mathop{\left\langle \vec{k},J=\frac{3}{2},J_{\tilde{z}}=\pm \frac{1}{2} \right|}}H'_2 \left| \vec{k},J=\frac{1}{2},J_{\tilde{z}}=\pm\frac{1}{2} \right\rangle^{\!\!(0)}  \nonumber \\
&=&  \alpha_K |\vec{E}\times \vec{k}| \left(\frac{-\sqrt{2}}{3}
\hbar \right).
\end{eqnarray}
All other matrix elements are zero.
Then one obtains
\begin{eqnarray}
&& \left|\vec{k},J=\frac{1}{2},J_{\tilde{z}}=\pm \frac{1}{2} \right\rangle^{\!\!(1)}
\label{eq:n-1-J=0.5-Jz=0.5} \\
&& =\left|\vec{k},J=\frac{3}{2},J_{\tilde{z}}=\pm \frac{1}{2} \right\rangle^{\!\!(0)}\frac{\alpha_K |\vec{E}\times \vec{k}| \frac{\sqrt{2}}{3} \hbar}{\Delta E}, \nonumber \\
&& \left|\vec{k},J=\frac{3}{2},J_{\tilde{z}}=\pm \frac{1}{2} \right\rangle^{\!\!(1)}
\label{eq:n-1-J=1.5-Jz=0.5} \\
&& = -\left|\vec{k},J=\frac{1}{2},J_{\tilde{z}}=\pm \frac{1}{2} \right \rangle^{\!\!(0)}\frac{\alpha_K |\vec{E}\times \vec{k}| \frac{\sqrt{2}}{3} \hbar}{\Delta E}, \nonumber \\
&& \left|\vec{k},J=\frac{3}{2},J_{\tilde{z}}=\pm \frac{3}{2} \right\rangle^{\!\!(1)} = 0.
\label{eq:n-1=J=1.5-Jz=1.5}
\end{eqnarray}
where $\Delta E \equiv E^{(0)}(\vec{k},J=3/2,J_{\tilde{z}}) -
E^{(0)}(\vec{k},J=1/2,J_{\tilde{z}})=3\alpha_{SO} \hbar^2/2$ is
independent of $\vec{k}$ and $J_{\tilde{z}}$.

Then $(j^{S}_{\alpha,\beta})_{sc}$ in Eq.~(\ref{eq:j-spin-im-1})
reduces to
\begin{eqnarray}
&& (j^{S}_{\alpha,\beta})_{sc}
\label{eq:j-conv-sc-contribution-formula} \\
&=& \mp \frac{1}{V}\frac{-e}{\hbar/2} \sum_{\vec{k}}\sum_{J_{\tilde z}=\pm 1/2} f^{(0)}\!\!\left(E^{(0)}(\vec{k},J)\right)
\nonumber \\
&& \times \frac{\alpha_K |\vec{E}\times \vec{k}| \frac{\sqrt{2}}{3} \hbar}{\Delta E}\left(
\sideset{^{(0)}\!\!}{}{\mathop{ \left\langle \vec{k},J',J_{\tilde z} \right|}} \! S_\alpha v^{(0)}_\beta \! \left| \vec{k},J,J_{\tilde z} \right\rangle^{\!\!(0)} \right. \nonumber \\
&& \ \ + \left.
\sideset{^{(0)}\!\!}{}{\mathop{ \left\langle \vec{k},J,J_{\tilde z} \right|}} S_\alpha v^{(0)}_\beta \left| \vec{k},J',J_{\tilde z} \right\rangle^{\!\!(0)} \right) , \nonumber
\end{eqnarray}
where the upper and lower signs apply to $J=3/2$ and $J=1/2$, respectively.
$J'=1/2$ ($3/2$) when $J=3/2$ ($1/2$).
Using the relation
\begin{eqnarray}
&& \sideset{^{(0)}\!\!}{}{\mathop{ \left\langle \vec{k},J\!=\!\frac{1}{2},J_{\tilde z} \right|}} \! S_\alpha v^{(0)}_\beta \! \left| \vec{k},J\!=\!\frac{3}{2},J_{\tilde z} \right\rangle^{\!\!(0)}  \\
&& = \frac{\sqrt{2}\hbar}{3}\frac{( \vec{E}\times \vec{k})_\alpha}{|\vec{E}\times \vec{k}|}\frac{\hbar k_\beta}{m}, \nonumber
\end{eqnarray}
one obtains
\begin{eqnarray}
&& (j^{S}_{\alpha,\beta})_{sc} \\
&=& \mp \frac{1}{V}\frac{-e}{\hbar/2} \sum_{\vec{k}}2 f^{(0)}\!\!\left(E^{(0)}(\vec{k},J)\right)
\frac{\alpha_K (\vec{E}\times \vec{k})_{\alpha} }{\Delta E}\frac{2\hbar^2}{9} \frac{2\hbar k_\beta}{m} \nonumber.
\end{eqnarray}
From the relation
\begin{eqnarray}
&& \frac{1}{V}\sum_{\vec{k}}f^{(0)}\!\!\left(E^{(0)}(\vec{k},J)\right) (\vec{E}\times \vec{k})_{\alpha} k_\beta \\
&=&  \frac{1}{V}\sum_{\vec{k}}f^{(0)}\!\!\left(E^{(0)}(\vec{k},J)\right)\left( - \frac{\epsilon_{\alpha\beta\gamma}}{3} E_\gamma\right) k^2 \nonumber \\
&=&  - \frac{\epsilon_{\alpha\beta\gamma}}{3} E_\gamma \frac{4\pi k_F^3/3}{(2\pi)^3}\frac{3k_F^2}{5},
\end{eqnarray}
one finally obtains
\begin{equation}
(j^{S}_{\alpha,\beta})_{sc} = \mp \frac{16}{135\pi^2}
\epsilon_{\alpha\beta\gamma}E_\gamma e \alpha_K k_F^3
\frac{\hbar^2 k_F^2 / 2m}{\Delta E}.
\end{equation}
Note that similarly to $(j^{S}_{\alpha,\beta})_{av}$,
$(j^{S}_{\alpha,\beta})_{sc}$ also has opposite signs for $J=3/2$
(upper sign) and $J=1/2$ (lower sign).

\subsection{Occupation change contribution}
\label{sec:j-spin-oc}
So far we have neglected impurity scattering.
Here we consider the scattering effect in the vanishing scattering strength limit.
Even in this limit, the scattering is important
since it violates the momentum conservation and allows electrons to relax in momentum space.
To illustrate its importance, it is useful to consider the case when the scattering is completely absent.
Then all throughout the adiabatic turning-on procedure of $H'_2$,
$\vec{k}$ remains a good quantum number
and the electron occupation in $\vec{k}$ space cannot be altered by $H'_2$,
which is in contrast to what we expect as illustrated in Fig.~2(b).
In the Kubo formalism, this effect is often addressed through the
vertex correction. Here we address this effect by noting that the
energy eigenvalues of $H_0+H'_2$ are bounded from below. In such a
situation, the electron occupation will relax in $\vec{k}$ space
to minimize the total energy of the electrons. Thus the occupation
change contribution $(j^{S}_{\alpha\beta})_{oc}$ to the spin
current density is given by
\begin{equation}
(j^{S}_{\alpha,\beta})_{oc}=\frac{1}{V}\frac{-e}{\hbar/2} {\rm
Tr}\left[  S_\alpha v^{(0)}_\beta \hat{\rho}^{(1)}_{oc} \right],
\label{eq:j-spin-conv-occupation-change}
\end{equation}
where $\hat{\rho}^{(1)}_{oc}$ denotes the first order correction to density matrix due to scattering
and is given by
\begin{equation}
\hat{\rho}^{(1)}_{oc} = \sum_{n}
f^{(1)}_n
\left|n\right\rangle^{\!(0)}
\sideset{^{(0)}\!}{}{\mathop{ \left\langle n \right|}}.
\label{eq:density-matrix-scattering}
\end{equation}
Here $f^{(1)}_n=f^{(0)}(E_n)-f^{(0)}(E^{(0)}_n)$ denotes the first order correction to the occupation function
and $E_n$ denotes the energy eigenvalue of $H_0+H'_2$.
$(j^{S}_{\alpha\beta})_{oc}$ is thus given by
\begin{equation}
(j^{S}_{\alpha,\beta})_{oc}=\frac{1}{V}\frac{-e}{\hbar/2} \sum_n
f^{(1)}\left(E_n\right) \sideset{^{(0)}\!}{}{\mathop{\left\langle
n \right|}}  S_\alpha v^{(0)}_\beta \left| n
\right\rangle^{\!(0)},
\end{equation}
where ${\rm Tr}[S_\alpha v^{(0)}_\beta \rho^{(0)}]=0$ has been used.

To determine $E_n$, it is useful to use the quantum numbers $\vec{k},J,J_{\tilde z}$ instead of $\vec{k},J,J_z$ to specify $n$
since for given $J$ sector, the state $|\vec{k},J,J_{\tilde z}\rangle^{\!(0)}$ diagonalizes $H_0+H'_2$
with the eigenvalue $E_n=E(\vec{k},J,J_{\tilde z})$ given by
\begin{equation}
E(\vec{k},J,J_{\tilde z})=E^{(0)}(\vec{k},J)+\alpha_K |\vec{E}\times \vec{k}| \frac{3\mp 1}{3} \hbar J_{\tilde z},
\end{equation}
where the upper and lower signs apply to $J=3/2$ and $J=1/2$, respectively.
To understand the effect of the second term, it is useful to consider one particular case;
$\vec{E}=E_z \hat{z}$.
Then the second term is proportional to $(k_x^2+k_y^2)^{1/2}$.
On the other hand, the first term is proportional to $\vec{k}^2=k_z^2+(k_x^2+k_y^2)$.
Thus the combined effect of the first and second terms is
to expand (shrink) the originally spherical Fermi surface along the ``equator" direction
when the second term is negative (positive).

The next step in the evaluation of $(j^{S}_{\alpha\beta})_{oc}$ is
to calculate $\sideset{^{(0)}\!}{}{\mathop{\left\langle n
\right|}}  S_\alpha v^{(0)}_\beta \left| n \right\rangle^{\!(0)}$
with $|n\rangle^{(0)}$ replaced by $|\vec{k},J,J_{\tilde
z}\rangle^{\!(0)}$. After straightforward calculation, one obtains
\begin{eqnarray}
&& \sideset{^{(0)}\!}{}{\mathop{\left\langle \vec{k},J,J_{\tilde z} \right|}}  S_\alpha v^{(0)}_\beta \left| \vec{k},J,J_{\tilde z} \right\rangle^{\!(0)} \\
&&=\frac{(\vec{E}\times \vec{k})_\alpha}{|\vec{E}\times \vec{k}|} \left( \pm \frac{\hbar J_{\tilde z}}{3} \right) \frac{\hbar k_\beta}{m}, \nonumber \\
&& = {\rm sgn}(E_z) \frac{k_x  \delta_{\alpha y} - k_y \delta_{\alpha x} }{\sqrt{k_x^2+k_y^2}}
\left( \pm \frac{\hbar J_{\tilde z}}{3} \right) \frac{\hbar k_\beta}{m}, \nonumber
\end{eqnarray}
where the upper and lower signs apply to $J=3/2$ and $J=1/2$, respectively.
After the average over the azimuthal angle in $\vec{k}$ space,
the above expression reduces to
\begin{eqnarray}
&& \overline{ \sideset{^{(0)}\!}{}{\mathop{\left\langle \vec{k},J,J_{\tilde z} \right|}}  S_\alpha v^{(0)}_\beta \left| \vec{k},J,J_{\tilde z} \right\rangle^{\!(0)} } \\
&& = {\rm sgn}(E_z) \frac{\sqrt{k_x^2+k_y^2}}{2} (-\epsilon_{\alpha\beta z}) \left( \pm \frac{\hbar J_{\tilde z}}{3} \right) \frac{\hbar }{m}, \nonumber
\end{eqnarray}

Then $(j^{S}_{\alpha\beta})_{oc}$ becomes
\begin{eqnarray}
&& (j^{S}_{\alpha,\beta})_{oc} \\
&=& \frac{1}{V}\frac{-e}{\hbar/2} \sum_{\vec{k}} \sum_{J_{\tilde z}}
f^{(0)}\!\!\left(E\left(\vec{k},J,J_{\tilde z} \right)\right) \nonumber \\
&& \times {\rm sgn}(E_z) \frac{\sqrt{k_x^2+k_y^2}}{2} (-\epsilon_{\alpha\beta z}) \left( \pm \frac{\hbar J_{\tilde z}}{3} \right) \frac{\hbar }{m}. \nonumber
\end{eqnarray}
After some tedious calculation, and for general direction of $\vec{E}$, one obtains
\begin{equation}
(j^{S}_{\alpha,\beta})_{oc} = \left\{
\begin{array}{c}
\displaystyle -10 \\
\displaystyle +2
\end{array} \right\} \times \frac{1}{27\pi^2}\epsilon_{\alpha\beta\gamma}E_\gamma e\alpha_K k_F^3.
\end{equation}
Here the upper and lower results apply to $J=3/2$ and $J=1/2$,
respectively. Note that $(j^{S}_{\alpha\beta})_{oc}$ has opposite
signs for $J=3/2$ and $J=1/2$.

\subsection{Summary}
Finally, $j^{S}_{\alpha,\beta}$ can be obtained by summing up all
three contributions, $(j^{S}_{\alpha\beta})_{oc}$,
$(j^{S}_{\alpha\beta})_{av}$, and $(j^{S}_{\alpha\beta})_{sc}$.
For $J=3/2$, one finds
\begin{equation}
j^{S}_{\alpha\beta}=-\frac{2}{9\pi^2}\epsilon_{\alpha\beta\gamma}
E_\gamma e \alpha_{K} k_F^3 \left( \frac{5}{3} - 1 + \frac{8}{15}
\frac{\hbar^{2}k_{F}^{2}/2m}{\Delta E}\right) \label{s24}
\end{equation}
and for $J=1/2$, one finds
\begin{equation}
j^{S}_{\alpha\beta}=\frac{2}{9\pi^2}\epsilon_{\alpha\beta\gamma}
E_\gamma e \alpha_{K} k_F^3 \left( \frac{1}{3} - 1 + \frac{8}{15}
\frac{\hbar^{2}k_{F}^{2}/2m}{\Delta E}\right) \label{s25}
\end{equation}

\section{Proper spin current density for $\vec{r}$}
\label{sec2}
In this section, we calculate the proper spin current density operator $j^{S,prop}_{\alpha,\beta}$
based on the conventional position operator $\vec{r}$. The corresponding operator $\hat{j}^{S,prop}_{\alpha,\beta}$ in Eq.~(\ref{eq:Proper-spin-current-operator}) may be divided into two pieces as follows,
\begin{equation}
\hat{j}^{S,prop}_{\alpha,\beta}=\hat{j}^{S}_{\alpha,\beta}+\hat{j}^{S,extra}_{\alpha,\beta},
\end{equation}
%
%
%
where $\hat{j}^{S}_{\alpha,\beta}$ is the conventional spin
current operator as defined in
Eq.~(\ref{eq:Conventional-spin-current-operator}), and
\begin{equation}
\hat{j}^{S,extra}_{\alpha,\beta}=\frac{1}{V}\frac{-e}{\hbar/2}
\frac{\left\{\frac{dS_\alpha}{dt},r_\beta
\right\}}{2}=\frac{1}{V}\frac{-e}{\hbar/2}\alpha_{SO}\epsilon_{\alpha\gamma\delta}
L_\gamma S_\delta r_\beta. \label{eq:Extra-spin-current-operator}
\end{equation}
%
Thus the difference $\hat{j}^{S,extra}_{\alpha,\beta}$ between
$j^{S,prop}_{\alpha,\beta}$ and $j^{S}_{\alpha,\beta}$ amounts to
the expectation value of $\hat{j}^{S,extra}_{\alpha,\beta}$,
\begin{equation}
j^{S,extra}_{\alpha,\beta}= {\rm Tr}\left[ \hat{j}^{S,extra}_{\alpha,\beta}
\hat{\rho} \right],
\label{eq:Extra-spin-current-density-formula}
\end{equation}
which will be evaluated below.
Among the two perturbations $H'_1$ and $H'_2$,
$H'_1$ cannot generate any contribution to $j^{S,extra}_{\alpha,\beta}$
since it does not induce any correlation between $\vec{r}$ (or $\vec{k}$)
and $\vec{L}$ (or $\vec{S}$).
Below we thus consider possible contribution from $H'_2$ only.

\subsection{Anomalous velocity contribution}
\label{sec:proper-anomalous}
By the ``anomalous velocity contribution", we refer to
$j^{S,extra}_{\alpha,\beta}$ with $\hat{\rho}$ in Eq.~(\ref{eq:Extra-spin-current-density-formula})
replaced by its equilibrium counterpart $\hat{\rho}^{(0)}$ in Eq.~(\ref{eq:Equilbrium-density-matrix}).
We find $j^{S,extra}_{\alpha,\beta}$ vanishes identically.
Below we demonstrate this for $\alpha=z$. The generalization to the case with $\alpha=x$ or $y$ is straightforward.
For $\alpha=z$, one obtains
\begin{eqnarray}
&& j^{S,extra}_{z,\beta}=\frac{1}{V}\frac{-e}{\hbar/2}\alpha_{SO} \sum_{\vec{k},J_z} f^{(0)}\left( E^{(0)}(\vec{k},J) \right) \\
&& \ \ \ \ \times
\sideset{^{(0)}\!\!}{}{\mathop{\left\langle \vec{k},J,J_z \right|}} (L_x S_y - L_y S_x)
r_\beta \left|\vec{k},J,J_z \right\rangle^{\!(0)}. \nonumber
\end{eqnarray}
To evaluate the expectation value in the above equation,
one uses the relations $[L_x S_y - L_y S_x,\vec{k}]=[L_x S_y - L_y S_x,J_z]=[r_\beta,J]=0$ to obtain
\begin{eqnarray}
&& \sideset{^{(0)}\!\!}{}{\mathop{\left\langle \vec{k},J,J_z \right|}} (L_x S_y - L_y S_x)
r_\beta \left|\vec{k},J,J_z \right\rangle^{\!(0)} \\
&=& \sideset{^{(0)}\!\!}{}{\mathop{\left\langle \vec{k},J,J_z \right|}} (L_x S_y - L_y S_x)
\left|\vec{k},J,J_z\right\rangle^{\!(0)} \nonumber \\
&& \times
\sideset{^{(0)}\!\!}{}{\mathop{\left\langle \vec{k},J,J_z\right|}} r_\beta \left|\vec{k},J,J_z \right\rangle^{(0)}. \nonumber
\end{eqnarray}
This expression vanishes since
\begin{equation}
\sideset{^{(0)}\!\!}{}{\mathop{\left\langle \vec{k},J,J_z \right|}} (L_x S_y - L_y S_x)
\left|\vec{k},J,J_z\right\rangle^{\!(0)}=0.
\end{equation}
This vanishing can be understood as follows.
Since $L_x S_y - L_y S_x$ is hermitian, its expectation value with respect to $|\vec{k},J,J_z\rangle^{\!(0)}$
must be real.
On the other hand, the conventional representations of $L_x S_y$ and $L_y S_x$ are pure imaginary.
The only way to reconcile these two properties is to make its expectation value zero.

Thus one finds $(j^{S,extra}_{\alpha,\beta})_{av}=0$ and
$(j^{S,prop}_{\alpha,\beta})_{av}=(j^{S}_{\alpha,\beta})_{av}$.
This way, one finally obtains
\begin{equation}
\left(j^{S,prop}_{\alpha,\beta}\right)_{av}= \pm \frac{2}{9\pi^2} \epsilon_{\alpha\beta\gamma}E_\gamma e\alpha_K k_F^3.
\label{eq:j-prop-anomalous-final-result}
\end{equation}
%

\subsection{State change contribution}
The state change contribution is defined as the contribution that arises from
the deviation of $\hat{\rho}$ from $\hat{\rho}^{(0)}$.
Thus the state change contribution from the extra spin current density operator is given by
\begin{equation}
\left({j}^{S,extra}_{\alpha,\beta}\right)_{sc}=\frac{1}{V}\frac{-e}{\hbar/2}\textup{Tr}\left[\frac{\left\{ \frac{dS_\alpha}{dt},r_\beta \right\}}{2} \hat{\rho}^{(1)} \right],
\end{equation}
where $\hat{\rho}^{(1)}$ is given in Eq.~(\ref{eq:density-matrix-no-scattering}).
Substituting $\hat{\rho}^{(1)}$ into the above equation leads to
\begin{eqnarray}
(j^{S,extra}_{\alpha,\beta})_{sc} \!\! &=&
\!\!\frac{1}{V}\frac{-e}{\hbar/2}  \sum_n \sum_{n'}
f^{(0)}\!\!\left(E^{(0)}_{n}\right) \alpha_{SO}
\epsilon_{\alpha\delta\gamma}
\\
&& \times \left(
\sideset{^{(0)}\!}{}{\mathop{ \left\langle n' \right|}}  L_\delta S_\gamma r_\beta \left| n \right\rangle^{\!(0)}
\sideset{^{(1)}\!}{}{\mathop{ \left\langle n \right|}} \left. n' \right\rangle^{\!(0)} \right. \nonumber \\
&& + \left.
\sideset{^{(0)}\!}{}{\mathop{ \left\langle n' \right|}} L_\delta S_\gamma r_\beta \left| n \right\rangle^{\!(1)}
\sideset{^{(0)}\!}{}{\mathop{ \left\langle n \right|}} \left. n' \right\rangle^{\!(0)} \right). \nonumber
\end{eqnarray}
Note that this expression has the identical structure as
Eq.~(\ref{eq:j-spin-im-1}) except that the conventional spin
current density operator $\hat{j}^{S}_{\alpha,\beta}$ is replaced
by the extra spin current density operator
$\hat{j}^{S,extra}_{\alpha,\beta}$. The evaluation of this
equation proceeds in a similar way. One first adopts the quantum
numbers $\vec{k}$, $J$, and $J_{\tilde z}$ to specify the state
$n$, where $J_{\tilde z}$ denotes the component of the total
angular momentum operator along $\vec{E}\times \vec{k}$ direction.
This allows one to utilize Eqs.~(\ref{eq:n-1-J=0.5-Jz=0.5}),
(\ref{eq:n-1-J=1.5-Jz=0.5}), and (\ref{eq:n-1=J=1.5-Jz=1.5}), and
one finds
\begin{eqnarray}
&& (j^{S,extra}_{\alpha,\beta})_{sc}
\\
&=& \mp\frac{1}{V}\frac{-e}{\hbar/2}  \sum_{\vec{k}} \sum_{J_{\tilde z} = \pm 1/2} f^{(0)}\!\!\left(E^{(0)}(\vec{k},J)\right)  \nonumber
\\
&& \times \alpha_{SO} \epsilon_{\alpha\delta\gamma}
\frac{\alpha_K\left| \vec{E}\times\vec{k}
\right|\frac{\sqrt{2}}{3}\hbar}{\Delta E}  \nonumber
\\
&& \times\left(
\sideset{^{(0)}\!}{}{\mathop{ \left\langle \vec{k},J',J_{\tilde z} \right|}} L_\delta S_\gamma r_\beta \left| \vec{k},J,J_{\tilde z} \right\rangle^{\!(0)}
 \right. \nonumber \\
&& + \left.
\sideset{^{(0)}\!}{}{\mathop{ \left\langle  \vec{k},J,J_{\tilde z} \right|}} L_\delta S_\gamma r_\beta \left|  \vec{k},J',J_{\tilde z} \right\rangle^{\!(0)}
\right) \nonumber
\\
&=& \mp\frac{1}{V}\frac{-e}{\hbar/2}  \sum_{\vec{k}} \sum_{J_{\tilde z} = \pm 1/2} f^{(0)}\!\!\left(E^{(0)}(\vec{k},J)\right) \nonumber
\\
&& \times \alpha_{SO}  \frac{\alpha_K\left| \vec{E}\times\vec{k}
\right|\frac{\sqrt{2}}{3}\hbar}{\Delta E} \nonumber
\\
&& \times  2\textup{Re}\left[
\sideset{^{(0)}\!}{}{\mathop{ \left\langle \vec{k},J',J_{\tilde z} \right|}}
\epsilon_{\alpha\delta\gamma} L_\delta S_\gamma r_\beta \left| \vec{k},J,J_{\tilde z} \right\rangle^{\!(0)}
 \right], \nonumber
\end{eqnarray}
where the upper and lower signs apply to $J=3/2$ and $J=1/2$, respectively. $J'=1/2\ (3/2)$ when $J'=3/2\ (1/2)$.
Using $[\vec{r},J]=[\vec{r},J_{\tilde z}]=[\vec{k},L_\delta]=[\vec{k},S_\gamma]=0$,
the last line of the above equation can be written as
\begin{eqnarray}
&& 2\textup{Re}\left[
\sideset{^{(0)}\!}{}{\mathop{ \left\langle \vec{k},J',J_{\tilde z} \right|}}
\epsilon_{\alpha\delta\gamma} L_\delta S_\gamma r_\beta \left| \vec{k},J,J_{\tilde z} \right\rangle^{\!(0)}\right] \ \ \ \ \ \ \ \ \ \ \\
&= &  2\textup{Re}\left[
\sideset{^{(0)}\!}{}{\mathop{ \left\langle \vec{k},J',J_{\tilde z} \right|}}
\epsilon_{\alpha\delta\gamma} L_\delta S_\gamma \left|\vec{k},J,J_{\tilde z}\right\rangle^{\!(0)} \right. \nonumber \\
&& \ \ \times \! \left.
\sideset{^{(0)}\!}{}{\mathop{\left\langle \vec{k},J,J_{\tilde z} \right|}}
r_\beta \left| \vec{k},J,J_{\tilde z} \right\rangle^{\!(0)}\right]. \nonumber
\end{eqnarray}
Here, $\sideset{^{(0)}\!}{}{\mathop{\left\langle
\vec{k},J,J_{\tilde z} \right|}} r_\beta \left|
\vec{k},J,J_{\tilde z} \right\rangle^{\!(0)}$ is manifestly real
since $r_\beta$ is hermitian. It can be also verified that
$\sideset{^{(0)}\!}{}{\mathop{ \left\langle \vec{k},J',J_{\tilde
z} \right|}} \epsilon_{\alpha\delta\gamma} L_\delta S_\gamma
\left|\vec{k},J,J_{\tilde z}\right\rangle^{\!(0)}$ is purely
imaginary. For this reason, the above equation vanishes
identically and one finds $(j^{S,extra}_{\alpha,\beta})_{sc} =0$.
Therefore
$(j^{S,prop}_{\alpha,\beta})_{sc}=(j^{S}_{\alpha,\beta})_{sc}$,
and one obtains
\begin{eqnarray}
(j^{S,prop}_{\alpha,\beta})_{sc}
&=& \mp \frac{16}{135\pi^2} \epsilon_{\alpha\beta\gamma}E_\gamma e \alpha_K k_F^3 \frac{\hbar^2 k_F^2 / 2m}{\Delta E}.  \nonumber
\end{eqnarray}
for $J=3/2$ (upper sign) and $J=1/2$ (lower sign), respectively.

\subsection{Occupation change contribution}
The occupation change contribution refers to the contribution arising from
the additional deviation of $\hat{\rho}$ from $\hat{\rho}^{(0)}$
due to the impurity scattering of infinitesimal strength.
Thus $(j^{S,extra}_{\alpha,\beta})_{oc}$ becomes
\begin{equation}
(j^{S,extra}_{\alpha,\beta})_{oc}= {\rm Tr}\left[  \hat{j}^{S,extra}_{\alpha,\beta} \hat{\rho}^{(1)}_{oc} \right],
\end{equation}
where $\hat{\rho}^{(1)}_{oc}$ denotes the impurity scattering effect to $\hat{\rho}$.
Using its expression in Eq.~(\ref{eq:density-matrix-scattering}), one obtains
\begin{eqnarray}
&& (j^{S,extra}_{\alpha,\beta})_{oc} \\
&&=\frac{1}{V}\frac{-e}{\hbar/2} \sum_n
f^{(1)}\left(E_n\right)\alpha_{SO}
\sideset{^{(0)}\!}{}{\mathop{\left\langle n \right|}}
\epsilon_{\alpha \gamma \delta} L_\gamma S_\delta r_\beta \left| n
\right\rangle^{\!(0)}. \nonumber
\end{eqnarray}
By following the same analysis as in Sec.~\ref{sec:proper-anomalous},
one can verify that
$\sideset{^{(0)}\!}{}{\mathop{\left\langle n \right|}} \epsilon_{\alpha \gamma \delta} L_\gamma S_\delta r_\beta \left| n \right\rangle^{\!(0)}=0$.
Thus $(j^{S,extra}_{\alpha,\beta})_{oc}$ vanishes identically and
$(j^{S,prop}_{\alpha,\beta})_{oc}=(j^{S}_{\alpha,\beta})_{oc}$.
Therefore one obtains
\begin{eqnarray}
(j^{S,prop}_{\alpha,\beta})_{oc}
&=& \left\{
\begin{array}{c}
\displaystyle -10 \\
\displaystyle +2
\end{array} \right\} \times \frac{1}{27\pi^2}\epsilon_{\alpha\beta\gamma}E_\gamma e\alpha_K k_F^3,  \nonumber
\end{eqnarray}
where the upper and lower numbers apply to $J=3/2$ and $J=1/2$, respectively.

\subsection{Summary}
In the preceding subsections, we showed that the extra spin
current density operator $\hat{j}^{S,extra}_{\alpha,\beta}$ does
not generate any extra contributions, so the proper spin current
density $j^{S,prop}_{\alpha,\beta}$ is identical to the
conventional spin current density $j^{S}_{\alpha,\beta}$. To
summarize the result of this section, we obtained
\begin{equation}
j^{S,prop}_{\alpha\beta}=-\frac{2}{9\pi^2}\epsilon_{\alpha\beta\gamma} E_\gamma e \alpha_{K} k_F^3
\left( \frac{5}{3} - 1 + \frac{8}{15} \frac{\hbar^{2}k_{F}^{2}/2m}{\Delta E}\right)
\end{equation}
for $J=3/2$, and
\begin{equation}
j^{S,prop}_{\alpha\beta}=\frac{2}{9\pi^2}\epsilon_{\alpha\beta\gamma} E_\gamma e \alpha_{K} k_F^3
\left( \frac{1}{3} - 1 + \frac{8}{15} \frac{\hbar^{2}k_{F}^{2}/2m}{\Delta E}\right)
\end{equation}
for $J=1/2$.

\section{Proper spin current density for $\vec{R}$}
\label{sec3}
In this section, we evaluate the proper spin current density
$j^{S,PROP}_{\alpha,\beta}$, which is the expectation value of the
proper spin current density operator
$\hat{j}^{S,PROP}_{\alpha,\beta}$
[Eq.~(\ref{eq:Consistent-spin-current-operator})] formulated in
terms of the proper position operator $\vec{R}$
[Eq.~(\ref{eq:Proper-position-operator})].
%
%
%
$\hat{j}^{S,PROP}_{\alpha,\beta}$ differs from the proper spin
current density operator $\hat{j}^{S,prop}_{\alpha,\beta}$ in that
the ``proper" position operator $\vec{R}$ in
Eq.~(\ref{eq:Proper-position-operator}) is used instead of the
conventional position operator $\vec{r}$. One way to understand
the difference between the two position operators is to compare
the corresponding velocity operators. The ``proper" velocity
operator $\vec{V}$ becomes
\begin{eqnarray}
\vec{V} &=& \frac{d\vec{R}}{dt}=\frac{[\vec{R},H_0+H'_1+H'_2]}{i\hbar} \\
&=& \frac{\hbar \vec{k}}{m}+\frac{\alpha_K}{\hbar}\vec{L}\times
\vec{E} \nonumber \\
&& +\frac{\alpha_K}{\hbar}\vec{L}\times \vec{E}
+\frac{\alpha_K^2}{e}\left( \vec{E}\times \vec{k} \right) \left(
\vec{k}\cdot \vec{L} \right) \nonumber \\
&& + \frac{\alpha_K \alpha_{SO}}{e}\vec{k}\times \left(
\vec{S}\times \vec{L} \right), \nonumber
\end{eqnarray}
where the first two terms amount to $\vec{v}$. Compared to the
conventional velocity operator $\vec{v}$ in
Eq.~(\ref{eq:Conventional-velocity-operator}), $\vec{V}$ differs
by
\begin{equation}
\vec{V}-\vec{v}=\delta \vec{v}^a +\delta \vec{v}^b + \delta \vec{v}^c,
\end{equation}
where
\begin{eqnarray}
\delta \vec{v}^a &=& \frac{\alpha_K}{\hbar}\vec{L}\times \vec{E}, \\
\delta \vec{v}^b &=& \frac{\alpha_K^2}{e}\left( \vec{E}\times \vec{k} \right) \left( \vec{k}\cdot \vec{L} \right), \\
\delta \vec{v}^c &=& \frac{\alpha_K \alpha_{SO}}{e}\vec{k}\times
\left( \vec{S}\times \vec{L} \right).
\end{eqnarray}
Among the three terms $\delta \vec{v}^a$, $\delta \vec{v}^b$, and
$\delta \vec{v}^c$, $\delta \vec{v}^a$ is identical to the
anomalous velocity $\vec{v}^{(1)}$ in
Eq.~(\ref{eq:Conventional-velocity-operator}). Recalling that
$\vec{v}^{(1)}$ is responsible for the anomalous velocity
contribution
$(j^{S}_{\alpha,\beta})_{av}=(j^{S,prop}_{\alpha,\beta})_{av}$ in
Eqs.~(\ref{eq:j-conv-anomalous-final-result}) and
(\ref{eq:j-prop-anomalous-final-result}), $\delta \vec{v}^a$ being
identical to $\vec{v}^{(1)}$ doubles the anomalous velocity
contribution, which will be verified explicitly in
Sec.~\ref{sec:consistant-anomalous}.
$\vec{v}^b$ is linear in $\vec{E}$ and thus generates a new piece
of the anomalous velocity. Compared to $\vec{v}^a$, it is smaller
by the dimensionless factor $\alpha_K \hbar k^2 / e$, which is
much smaller than 1 in the small $\vec{k}$ limit. Thus $\vec{v}^b$
is not important in the small $\vec{k}$ limit, but just for the
sake of completeness, we evaluate its contribution to
$j^{S,PROP}_{\alpha,\beta}$ in
Sec.~\ref{sec:consistant-anomalous}. On the other hand,
$\vec{v}^c$ is zeroth order in $\vec{E}$. We examine its possible
contribution below.

For explicit evaluation of $j^{S,PROP}_{\alpha,\beta}$, one needs
to deal with $\hat{j}^{S,PROP}_{\alpha,\beta}$ in
Eq.~(\ref{eq:Consistent-spin-current-operator}). It is useful to
compare $\hat{j}^{S,PROP}_{\alpha,\beta}$ with
$\hat{j}^{S}_{\alpha,\beta}$ and
$\hat{j}^{S,prop}_{\alpha,\beta}$,
\begin{eqnarray}
\hat{j}^{S,PROP}_{\alpha,\beta}&=&\hat{j}^{S,prop}_{\alpha,\beta}+\hat{j}^{S,EXTRA}_{\alpha,\beta}
\label{eq:Consistent-vs-Proper} \\
&=& \hat{j}^{S}_{\alpha,\beta}+\hat{j}^{S,extra}_{\alpha,\beta}+
\hat{j}^{S,EXTRA}_{\alpha,\beta}, \nonumber
\end{eqnarray}
where $\hat{j}^{S,extra}_{\alpha,\beta}$ is defined in
Eq.~(\ref{eq:Extra-spin-current-operator}) and
$\hat{j}^{S,EXTRA}_{\alpha,\beta}$ is given by
\begin{equation}
\hat{j}^{S,EXTRA}_{\alpha,\beta}=\hat{j}^{S,a}_{\alpha,\beta}+\hat{j}^{S,b}_{\alpha,\beta}+\hat{j}^{S,c}_{\alpha,\beta}.
\label{eq:j-EXTRA-three-terms}
\end{equation}
Here
\begin{eqnarray}
\hat{j}^{S,a}_{\alpha,\beta} \!\! & = & \!\! \frac{1}{V}\frac{-e}{\hbar/2} \frac{\alpha_K}{\hbar}S_\alpha \left( \vec{L}\times \vec{E} \right)_\beta, \\
\hat{j}^{S,b}_{\alpha,\beta} \!\! & = & \!\! \frac{1}{V}\frac{-e}{\hbar/2} \frac{\alpha_K^2}{e} S_\alpha \left( \vec{E} \times \vec{k} \right)_\beta \left(\vec{k}\cdot \vec{L} \right), \\
\hat{j}^{S,c}_{\alpha,\beta} \!\! & = & \!\!
\frac{1}{V}\frac{-e}{\hbar/2} \frac{\alpha_K \alpha_{SO}}{2e}
\left[ \frac{\hbar^2}{2} \left( \delta_{\alpha\beta}\vec{k}\cdot\vec{L}
- k_\alpha L_\beta \right) \right. \label{eq:j-S-extra-c} \\
&& \ \ \ \ \ - \left. \left\{ \left( \vec{S}\times \vec{L} \right)_\alpha, \left( \vec{k}\times \vec{L} \right)_\beta\right\} \right]. \nonumber
\end{eqnarray}
It is evident that the expectation value of
$\hat{j}^{S,EXTRA}_{\alpha,\beta}$ determines the difference
between $j^{S,PROP}_{\alpha,\beta}$ and the two former spin
current densities $j^{S}_{\alpha,\beta}$ and
$j^{S,prop}_{\alpha,\beta}$.

Simple order counting helps estimate effects of the three terms of $\hat{j}^{S,EXTRA}_{\alpha,\beta}$.
Since $\hat{j}^{S,a}$ and $\hat{j}^{S,b}$ are linear in $\vec{E}$,
they can affect only the anomalous velocity contribution.
They do not affect the state change contribution and the occupation change contribution.
Among these terms $\hat{j}^{S,b}_{\alpha,\beta}$ is smaller than $\hat{j}^{S,a}_{\alpha,\beta}$
by $\alpha_K \hbar k^2 / e$, which approaches zero in the small $\vec{k}$ limit.
Thus $\hat{j}^{S,a}_{\alpha,\beta}$ is expected to be more important.
Actually it can be easily verified that $\hat{j}^{S,a}_{\alpha,\beta}$ is identical to
$\frac{1}{V}\frac{-e}{\hbar/2}S_\alpha v^{(1)}_\beta$ [see Eq.~(\ref{eq:j-spin-v1})],
which is responsible for the anomalous velocity contribution of $j^{S}_{\alpha,\beta}$.
Thus the presence of $\hat{j}^{S,a}_{\alpha,\beta}$ doubles the anomalous velocity contribution.
On the other hand, $\hat{j}^{S,c}_{\alpha,\beta}$ differs by the factor $\alpha_K \alpha_{SO} \hbar m /e$
from $\frac{1}{V}\frac{-e}{\hbar/2}S_\alpha v^{(0)}_\beta$ [see Eq.~(\ref{eq:j-spin-conv-occupation-change})],
which is responsible for the occupation change contribution of $j^{S}_{\alpha,\beta}$.
Since this factor may not be small in the strong spin-orbit coupling limit that we consider,
it is yet unclear how important $\hat{j}^{S,c}_{\alpha,\beta}$ is.

Below we demonstrate through explicit calculation that
$\hat{j}^{S,c}_{\alpha,\beta}$ does not generate any important contribution
and the only important effect of $\hat{j}^{S,EXTRA}_{\alpha,\beta}$ is
to double the anomalous velocity contribution.

\subsection{Anomalous velocity contribution}
\label{sec:consistant-anomalous}
To calculate the anomalous velocity contribution
$(j^{S,PROP}_{\alpha,\beta})_{av}$ to $j^{S,PROP}_{\alpha,\beta}$,
it is sufficient to evaluate $(j^{S,EXTRA}_{\alpha,\beta})_{av}$,
which is given by
\begin{eqnarray}
(j^{S,EXTRA}_{\alpha,\beta})_{av} &=& {\rm Tr}\left[ \hat{j}^{S,EXTRA}_{\alpha,\beta}\hat{\rho}^{(0)}\right] \\
&=& (j^{S,a}_{\alpha,\beta})_{av}+(j^{S,b}_{\alpha,\beta})_{av}+(j^{S,c}_{\alpha,\beta})_{av}, \nonumber
\end{eqnarray}
where
\begin{eqnarray}
(j^{S,a}_{\alpha,\beta})_{av} &=& {\rm Tr}\left[ \hat{j}^{S,a}_{\alpha,\beta}\hat{\rho}^{(0)}\right],  \\
(j^{S,b}_{\alpha,\beta})_{av} &=& {\rm Tr}\left[ \hat{j}^{S,b}_{\alpha,\beta}\hat{\rho}^{(0)}\right], \\
(j^{S,c}_{\alpha,\beta})_{av} &=& {\rm Tr}\left[ \hat{j}^{S,c}_{\alpha,\beta}\hat{\rho}^{(0)}\right].
\end{eqnarray}

For $(j^{S,c}_{\alpha,\beta})_{av}$, it vanishes simply because
$\hat{j}^{S,c}_{\alpha,\beta}$ is linear in $\vec{k}$ whereas $\hat{\rho}^{(0)}$ puts the same weighting independent of the direction of $\vec{k}$.
To evaluate $(j^{S,a}_{\alpha,\beta})_{av}$, one notes
\begin{equation}
\hat{j}^{S,a}_{\alpha,\beta} = \frac{1}{V}\frac{-e}{\hbar/2} S_\alpha \delta v^{a}_\beta
=\frac{1}{V}\frac{-e}{\hbar/2} S_\alpha v^{(1)}_\beta.
\end{equation}
Thus $(j^{S,a}_{\alpha,\beta})_{av}$ is identical to
$(j^{S}_{\alpha,\beta})_{av}$ in Eq.~(\ref{eq:j-spin-av}), and one
obtains
\begin{equation}
\left(j^{S,a}_{\alpha,\beta}\right)_{av}= \pm \frac{2}{9\pi^2} \epsilon_{\alpha\beta\gamma}E_\gamma e\alpha_K k_F^3,
\end{equation}
for $J=3/2$ (upper sign) and $J=1/2$ (lower sign).

To evaluate $(j^{S,b}_{\alpha,\beta})_{av}$, one notes
\begin{equation}
\hat{j}^{S,b}_{\alpha,\beta} = \frac{1}{V}\frac{-e}{\hbar/2} S_\alpha \delta v^{b}_\beta.
\end{equation}
We demonstrate the evaluation of $(j^{S,b}_{\alpha,\beta})_{av}$ for $\vec{E}=E_z \hat{z}$.
In this case, it is straightforward to verify that $(j^{S,b}_{\alpha,\beta})_{av}$ is proportional to $\epsilon_{\alpha\beta z}$.
It then suffices to evaluate $\epsilon_{\alpha\beta z}(j^{S,b}_{\alpha,\beta})_{av}$.
One uses the following relation
\begin{eqnarray}
&& \epsilon_{\alpha\beta z}S_\alpha \left( \vec{E} \times \vec{k} \right)_\beta \left( \vec{k} \cdot \vec{L} \right) \\
 &=& \left( S_x k_x + S_y k_y \right) \left( \vec{k} \cdot \vec{L} \right) E_z \nonumber
\end{eqnarray}
to obtain
\begin{eqnarray}
&& \epsilon_{\alpha \beta z} \left(j^{S,b}_{\alpha,\beta}\right)_{av} \\
&=&
\frac{1}{V}\frac{-e}{\hbar/2} \frac{\alpha_K^2}{e} E_z {\rm Tr} \left[ \left( S_x k_x + S_y k_y \right) \left( \vec{k} \cdot \vec{L} \right)  \hat{\rho}^{(0)}\right]. \nonumber
\end{eqnarray}
Since $f^{(0)}(E^{(0)}_n)$ in $\hat{\rho}^{(0)}$
[Eq.~(\ref{eq:Equilbrium-density-matrix})] does not depend on the
direction of $\vec{k}$, this expression may survive only when the
traced expression is even in components of $\vec{k}$. It thus
reduces to
\begin{eqnarray}
&&\epsilon_{\alpha \beta z} \left(j^{S,b}_{\alpha,\beta}\right)_{av} \\
&=&
\frac{1}{V}\frac{-e}{\hbar/2} \frac{\alpha_K^2}{e} E_z{\rm Tr} \left[ \left( S_x L_x k_x^2 + S_y L_y k_y^2 \right) \hat{\rho}^{(0)}\right]. \nonumber
\end{eqnarray}
Since the dependence on $\vec{k}$ is decoupled from the dependencies on $\vec{S}$ and $\vec{L}$ as far as $\hat{\rho}^{(0)}$ is concerned,
$k_x^2$ and $k_y^2$ in the above expression may be replaced by $k^2/3$,
and $S_x L_x$ and $S_y L_y$ by $\vec{S}\cdot \vec{L} /3 $.
Thus the above expression reduces further to
\begin{eqnarray}
&&\epsilon_{\alpha \beta z} \left(j^{S,b}_{\alpha,\beta}\right)_{av} \\
&=&
\frac{1}{V}\frac{-e}{\hbar/2} \frac{\alpha_K^2}{e} E_z \frac{2}{9} {\rm Tr} \left[ \left( \vec{S}\cdot \vec{L} \right) k^2 \hat{\rho}^{(0)}\right]. \nonumber
\end{eqnarray}
Here $\vec{S}\cdot \vec{L}$ may be replaced by $(\hbar^2/2)[ J(J+1)-1\cdot 2-\frac{1}{2}\frac{3}{2}]$,
which is $\hbar^2/2$ for $J=3/2$ and $-\hbar^2$ for $J=1/2$.
The remaining calculation is straightforward. Generalizing to general direction of $\vec{E}$, one obtains
\begin{equation}
\left(j^{S,b}_{\alpha,\beta}\right)_{av}= \mp \frac{1}{15\pi^2}
\epsilon_{\alpha\beta\gamma}E_\gamma \hbar \alpha_K^2 k_F^5
\end{equation}
for $J=3/2$ (upper sign) and $J=1/2$ (lower sign).
Note that $(j^{S,b}_{\alpha,\beta})_{av}$ differs from $(j^{S,a}_{\alpha,\beta})_{av}$
by the factor $-\frac{3}{5}\alpha_K \hbar k_F^2 /e$, which is smaller than 1 in the small $\vec{k}$ limit.

Therefore $(j^{S,EXTRA}_{\alpha,\beta})_{av}$ is given by
\begin{equation}
\left(j^{S,EXTRA}_{\alpha,\beta}\right)_{av}= \pm \frac{2}{9\pi^2}
\epsilon_{\alpha\beta\gamma}E_\gamma e\alpha_K k_F^3 \left( 1
-\frac{3}{10} \frac{\alpha_K \hbar k_F^2}{e} \right).
\end{equation}
Considering the relation between
$\hat{j}^{S,EXTRA}_{\alpha,\beta}$ and
$\hat{j}^{S,PROP}_{\alpha,\beta}$ in
Eq.~(\ref{eq:Consistent-vs-Proper}), one finally obtains
\begin{equation}
\left(j^{S,PROP}_{\alpha,\beta}\right)_{av}= \pm \frac{4}{9\pi^2}
\epsilon_{\alpha\beta\gamma}E_\gamma e\alpha_K k_F^3 \left( 1
-\frac{3}{20} \frac{\alpha_K \hbar k_F^2}{e} \right),
\end{equation}
for $J=3/2$ (upper sign) and $J=1/2$ (lower sign).

\subsection{State change contribution}
The perturbation $H'_1$ and $H'_2$ can modify the density matrix
$\hat{\rho}$. To calculate the state change contribution
$(j^{S,PROP}_{\alpha,\beta})_{sc}$ arising from the density matrix
change (in the absence of any scattering), it is sufficient to
retain only $\hat{j}^{S,c}_{\alpha,\beta}$ out of the three terms
for $\hat{j}^{S,EXTRA}_{\alpha,\beta}$
[Eq.~(\ref{eq:j-EXTRA-three-terms})] and calculate
\begin{equation}
(j^{S,c}_{\alpha,\beta})_{sc}={\rm Tr}\left[
\hat{j}^{S,c}_{\alpha,\beta}\hat{\rho}^{(1)}\right],
\end{equation}
where $\hat{\rho}^{(1)}$ denotes the first order (in $\vec{E}$)
change of $\hat{\rho}$. Here we may ignore the contributions from
$\hat{j}^{S,a}_{\alpha,\beta}$ and $\hat{j}^{S,b}_{\alpha,\beta}$,
since these operators are already first order in $\vec{E}$ and
generate the second order contribution when combined with
$\hat{\rho}^{(1)}$.

Both $H'_1$ and $H'_2$ contribute to $\hat{\rho}^{(1)}$. However
$\hat{\rho}^{(1)}$ due to $H'_1$ does not contribute to
$(j^{S,c}_{\alpha,\beta})_{sc}$ since as far as $H_0+H'_1$ is
concerned, there is no coupling between ($\vec{k}$, $\vec{r}$) and
($\vec{L}$, $\vec{S}$). Combined with the fact that
$\hat{j}^{S,c}_{\alpha,\beta}$ is odd in $\vec{L}$ or $\vec{S}$,
this feature prohibits $H'_1$ from generating any contribution to
$(j^{S,c}_{\alpha,\beta})_{sc}$.

Below we confine ourselves to $\hat{\rho}^{(1)}$ arising from
$H'_2$, which has been explicitly constructed in
Sec.~\ref{sec:j-spin-im}. Following the similar calculation
procedure in Sec.~\ref{sec:j-spin-im}, one obtains
\begin{eqnarray}
&& (j^{S,c}_{\alpha,\beta})_{sc}
\label{eq:j-spin-S-c-evaluation} \\
&=& \mp \sum_{\vec{k}}\sum_{J_{\tilde z}=\pm 1/2}
f^{(0)}\!\!\left(E^{(0)}(\vec{k},J)\right)
\nonumber \\
&& \times \frac{\alpha_K |\vec{E}\times \vec{k}|
\frac{\sqrt{2}}{3} \hbar}{\Delta E} \nonumber \\
&&  \times 2 {\rm Re}\left\{  \sideset{^{(0)}\!\!}{}{\mathop{
\left\langle \vec{k},J',J_{\tilde z} \right|}}
\hat{j}^{S,c}_{\alpha,\beta} \left| \vec{k},J,J_{\tilde z}
\right\rangle^{\!\!(0)} \right\} , \nonumber
\end{eqnarray}
where the upper and lower signs apply to $J=3/2$ and $J=1/2$,
respectively. $J'=1/2$ ($3/2$) when $J=3/2$ ($1/2$). This is the
counterpart of Eq.~(\ref{eq:j-conv-sc-contribution-formula}).
From symmetry consideration, it can be verified that $(j^{S,c}_{\alpha,\beta})_{sc}$ should be
proportional to $\epsilon_{\alpha\beta \gamma}E_\gamma$.
Also the expression for $\hat{j}^{S,c}_{\alpha,\beta}$ in Eq.~(\ref{eq:j-S-extra-c}) indicates that
$(j^{S,c}_{\alpha,\beta})_{sc}$ in Eq.~(\ref{eq:j-spin-S-c-evaluation}) is of the order of $\vec{E}\hbar \alpha_K k_F^5$.
Thus one finds
\begin{equation}
(j^{S,c}_{\alpha,\beta})_{sc}=\eta \epsilon_{\alpha\beta\gamma}E_\gamma \hbar \alpha_K^2 k_F^5,
\end{equation}
where $\eta$ is a dimensionless constant.
Explicit evaluation of Eq.~(\ref{eq:j-spin-S-c-evaluation}) is necessary to determine $\eta$.
However even without the explicit evaluation, it is evident that
$(j^{S,c}_{\alpha,\beta})_{sc}$ is smaller than $(j^{S,PROP}_{\alpha,\beta})_{av}$ by
the factor $\alpha_K \hbar k_F^2/e$, which is smaller than 1 in the smaller $\vec{k}$ limit.
Therefore in the small $\vec{k}$ limit, $(j^{S,c}_{\alpha,\beta})_{sc}$ is not important.

Below we demonstrate the explicit evaluate of $(j^{S,c}_{\alpha,\beta})_{sc}$ to determine $\eta$.
It suffices to assume $\vec{E}=E_z\hat{z}$ and
evaluate $\epsilon_{\alpha\beta z}(j^{S,c}_{\alpha,\beta})_{sc}$.
For this, one utilizes Eq.~(\ref{eq:j-S-extra-c}) to obtain
\begin{eqnarray}
\epsilon_{\alpha\beta z}\hat{j}^{S,c}_{\alpha,\beta}  &=&
\frac{1}{V} \frac{-e}{\hbar/2} \frac{\alpha_K \alpha_{SO}}{2e}
\left[ -\frac{\hbar^2}{2} \left(\vec{k}\times \vec{L}\right)_z
\right.
\label{eq:j-extra-c-antisymmetric-component} \\
&& \left. \rule{0mm}{5mm} -L_z
(\vec{S}\cdot\vec{k}\times\vec{L})-(\vec{k}\times\vec{L}\cdot\vec{S})L_z
\right]. \nonumber
\end{eqnarray}
After some algebra, one finds
\begin{eqnarray}
&& \epsilon_{\alpha\beta z}2{\rm Re}
\left\{  \sideset{^{(0)}\!\!}{}{\mathop{
\left\langle \vec{k},J',J_{\tilde z} \right|}}
\hat{j}^{S,c}_{\alpha,\beta} \left| \vec{k},J,J_{\tilde z}
\right\rangle^{\!\!(0)} \right\} \\
&& = \frac{1}{V} \frac{-e}{\hbar/2} \frac{\alpha_K \alpha_{SO}}{2e}
\frac{3\sqrt{2}}{2}\hbar^3 \left|\hat{z}\times \vec{k}\right|
\end{eqnarray}
The rest of calculation is straightforward and results in
\begin{equation}
(j^{S,c}_{\alpha,\beta})_{sc}=\pm \frac{2}{45\pi^2} \epsilon_{\alpha\beta\gamma}E_\gamma \hbar \alpha_K^2 k_F^5.
\end{equation}
Finally by combining with $(j^{S,prop}_{\alpha,\beta})_{sc}$,
one obtains
\begin{eqnarray}
&& (j^{S,PROP}_{\alpha,\beta})_{sc} \\
&=& \mp \frac{2}{45\pi^2} \epsilon_{\alpha\beta\gamma}E_\gamma e \alpha_K k_F^3
\left( \frac{8}{3} \frac{\hbar^2 k_F^2/2m}{\Delta E} - \frac{\alpha_K \hbar k_F^2}{e} \right), \nonumber
\end{eqnarray}
for $J=3/2$ (upper sign) and $J=1/2$ (lower sign).

\subsection{Occupation change contribution}
The occupation change contribution refers to the contribution arising from
the additional deviation of $\hat{\rho}$ from $\hat{\rho}^{(0)}$
due to the impurity scattering of infinitesimal strength.
Due to Eqs.~(\ref{eq:Consistent-vs-Proper}) and (\ref{eq:j-EXTRA-three-terms}),
the calculation of $(j^{S,PROP}_{\alpha,\beta})_{oc}$ overlaps a lot
with that of $(j^{S,prop}_{\alpha,\beta})_{oc}$ and $(j^{S}_{\alpha,\beta})_{oc}$.
The only piece that requires additional calculation is $(j^{S,c}_{\alpha,\beta})_{oc}$, which is given by
\begin{equation}
(j^{S,c}_{\alpha,\beta})_{oc}= {\rm Tr}\left[  \hat{j}^{S,c}_{\alpha,\beta} \hat{\rho}^{(1)}_{oc} \right],
\end{equation}
where $\hat{\rho}^{(1)}_{oc}$ denotes the impurity scattering effect to $\hat{\rho}$.
The perturbation $H'_1$ does not make any contribution to $(j^{S,c}_{\alpha,\beta})_{oc}$
since it does not induce any correlation among $(\vec{k},\vec{r})$ and $(\vec{L},\vec{S})$.
Below we thus consider the perturbation $H'_2$ only.
Then using the expression for $\hat{\rho}^{(1)}_{oc}$ in Eq.~(\ref{eq:density-matrix-scattering}), one obtains
\begin{equation}
(j^{S,c}_{\alpha,\beta})_{oc} = \sum_n
f^{(1)}\left(E_n\right)
\sideset{^{(0)}\!}{}{\mathop{\left\langle n \right|}}
\hat{j}^{S,c}_{\alpha,\beta} \left| n
\right\rangle^{\!(0)}.
\end{equation}

From symmetry consideration, one can verify that $(j^{S,c}_{\alpha,\beta})_{oc}$ should be proportional to $\epsilon_{\alpha\beta\gamma}E_\gamma$.
It then suffices to assume $\vec{E}=E_z \hat{z}$ and evaluate $\epsilon_{\alpha\beta z}(j^{S,c}_{\alpha,\beta})_{oc}$.
For its evaluation, one uses Eq.~(\ref{eq:j-extra-c-antisymmetric-component})
and also the relation,
\begin{eqnarray}
&& \sum_{k_z} \left\{ \sideset{^{(0)}\!}{}{\mathop{\left\langle \vec{k},J,J_{\tilde z} \right|}} \frac{\hbar^2}{2}\left(\vec{k}\times \vec{L} \right)
\left| \vec{k},J,J_{\tilde z} \right\rangle^{\!(0)} \right. \\
&& +  \sideset{^{(0)}\!}{}{\mathop{\left\langle \vec{k},J,J_{\tilde z} \right|}}
L_z \left( \vec{S} \cdot \vec{k}\times \vec{L} \right)
\left| \vec{k},J,J_{\tilde z} \right\rangle^{\!(0)}    \nonumber \\
&& + \left.  \sideset{^{(0)}\!}{}{\mathop{\left\langle \vec{k},J,J_{\tilde z} \right|}}
\left(\vec{k}\times \vec{L} \cdot \vec{S} \right) L_z
\left| \vec{k},J,J_{\tilde z} \right\rangle^{\!(0)}  \right\} =0, \nonumber
\end{eqnarray}
which shows that $(j^{S,c}_{\alpha,\beta})_{oc}=0$.
Finally by combining with $(j^{S,prop}_{\alpha,\beta})_{oc}$, one obtains
\begin{eqnarray}
(j^{S,PROP}_{\alpha,\beta})_{oc}
&=& \left\{
\begin{array}{c}
\displaystyle -10 \\
\displaystyle +2
\end{array} \right\} \times \frac{1}{27\pi^2}\epsilon_{\alpha\beta\gamma}E_\gamma e\alpha_K k_F^3,  \nonumber
\end{eqnarray}
where the upper and lower results apply to $J=3/2$ and $J=1/2$, respectively.

\subsection{Summary}
To summarize the result of this section, we obtained
\begin{eqnarray}
j^{S,PROP}_{\alpha\beta} &=& -\frac{2}{9\pi^2}\epsilon_{\alpha\beta\gamma} E_\gamma e \alpha_{K} k_F^3
 \\
&& \times \left[ \frac{5}{3} -  \left( 2 - \frac{3}{10}\frac{\alpha_K \hbar k_F^2}{e} \right) \right. \nonumber \\
&& \ \ \ \ + \left. \left( \frac{8}{15} \frac{\hbar^{2}k_{F}^{2}/2m}{\Delta E} -\frac{1}{5}\frac{\alpha_K \hbar k_F^2}{e} \right) \right] \nonumber
\end{eqnarray}
for $J=3/2$ and
\begin{eqnarray}
j^{S,PROP}_{\alpha\beta} &=& \frac{2}{9\pi^2}\epsilon_{\alpha\beta\gamma} E_\gamma e \alpha_{K} k_F^3
 \\
&& \times \left[ \frac{1}{3} -  \left( 2 - \frac{3}{10}\frac{\alpha_K \hbar k_F^2}{e} \right) \right. \nonumber \\
&& \ \ \ \ + \left. \left( \frac{8}{15} \frac{\hbar^{2}k_{F}^{2}/2m}{\Delta E} -\frac{1}{5}\frac{\alpha_K \hbar k_F^2}{e} \right) \right] \nonumber
\end{eqnarray}
for $J=1/2$.
Note that $j^{S,PROP}_{\alpha\beta}$ differs from $j^{S}_{\alpha\beta}$ and $j^{S,prop}_{\alpha\beta}$ in two ways.
One difference is the extra terms, which are of order of $\vec{E}\alpha_K^2 \hbar k_F^5$
and thus smaller than other leading order terms by the factor $\alpha_K\hbar k_F^2/e$.
Since this factor approaches zero in the small $\vec{k}$ regime that we consider,
this difference is not important.
The other difference is the factor two enhancement of the anomalous velocity contribution.
Since this enhancement occurs at the leading order term of the order of $\vec{E}e\alpha_K k_F^3$,
this enhancement by factor 2 is relevant.
Thus the only important deviation
of $j^{S,PROP}_{\alpha\beta}$ from $j^{S}_{\alpha\beta}$ and $j^{S,prop}_{\alpha\beta}$
is the factor two enhancement of the anomalous velocity contribution.

